\documentclass[longauth]{aa}
\usepackage{txfonts}
\usepackage{graphicx}
\usepackage{natbib}
\usepackage{subeqnarray}
\usepackage{cases}
\usepackage{ulem}
\usepackage{rotating}
\usepackage{makecell}

\newcommand{\co}{$^{12}$CO}       
\newcommand{\xco}{$^{13}$CO}       
\newcommand{\yco}{C$^{18}$O}      
\newcommand{\hco}{HCO$^{+}$}       
\newcommand{\NH}{NH$_{3}$}       
    
\newcommand{\jxs}{($J = 1 - 0$) }  
\newcommand{\kms}{km s$^{-1}$}     
\newcommand{\msun}{M$_{\rm \odot}$}

\newcommand{\cmm}{cm$^{-2}$}

\usepackage[colorlinks=true,citecolor=blue]{hyperref}
\begin{document}

\title{Cloud-cloud collision and star formation in G013.313+0.193}
\author{Dilda Berdikhan \inst{1,2,3} 
\and Jarken Esimbek \inst{1,2.4,5}
\and Christian Henkel \inst{6,1,7}
\and Ye Xu \inst{8}
\and Jianjun Zhou \inst{1,2,4,5}
\and De-Jian Liu \inst{9}
\and Ernazar Abdikamalov \inst{10,3}
\and Yingxiu Ma \inst{1,4,5}
\and Toktarkhan Komesh \inst{3,11}
\and Yuxin He \inst{1,2,4,5}
\and Wenjun Zhang \inst{1,2}
\and Xindi Tang \inst{1,2,4,5}
\and Gang Wu \inst{1,6}
\and Dalei Li \inst{1,2,4,5}
\and Dongdong Zhou \inst{1,4,5}
\and Kadirya Tursun \inst{1,4,5}
\and Hailiang Shen \inst{1,2}
\and Ernar Imanaly \inst{1,2}
\and Qaynar Jandaolet \inst{1,2}
\and Arailym Manapbayeva \inst{11}
\and Duriya Tuiakbayeva \inst{11,12}
}

\institute{
Xinjiang Astronomical Observatory, Chinese Academy of Sciences, 830011 Urumqi, P. R. China \\
e-mail: dilda@xao.ac.cn, jarken@xao.ac.cn, chenkel@mpifr-bonn.mpg.de
\and 
University of Chinese Academy of Sciences, 100080 Beijing, P. R. China 
\and
Energetic Cosmos Laboratory, Nazarbayev University, Astana 010000, Kazakhstan
\and 
State Key Laboratory of Radio Astronomy and Technology, A20 Datun Road, Chaoyang District, Beijing, 100101, P. R. China
\and
Xinjiang Key Laboratory of Radio Astrophysics, Urumqi 830011, P. R. China
\and
Max-Planck-Institut f\"ur Radioastronomie, Auf dem H\"ugel 69, 53121 Bonn, Germany 
\and
Astronomy Department, King Abdulaziz University, PO Box 80203, 21589 Jeddah, Saudi Arabia
\and
Purple Mountain Observatory, Chinese Academy of Sciences, Nanjing 210023, P. R. China
\and 
College of Mathematical and Physics, China Three Corges University,Yichang 443002, P. R. China
\and 
Department of Physics, Nazarbayev University, Astana 010000, Kazakhstan
\and
Institute of Experimental and Theoretical Physics, Al-Farabi Kazakh National University, Almaty 050040, Kazakhstan
\and 
Sh. Ualikhanov Kokshetau University, 76 Abaya St., 020000, Kokshetau, Kazakhstan
}

\abstract
{
We study the G013.313+0.193 (G013.313) region, a complex environment characterised by molecular cloud interactions indicative of cloud-cloud collision (CCC). Observations of the \NH\,(1,1) and (2,2) inversion transitions were obtained using the Nanshan 26 m radio telescope, while \hco~\jxs, \co, \xco, and \yco~\jxs transitions from the 14 m Purple Mountain Observatory Delingha (PMODLH) 14 m telescope. Archival data are also included. We identified key observational signatures of CCC, including complementary spatial distributions, U-shaped structures, bridge features, and V-shaped velocity distributions. The position-velocity (P-V) diagrams reveal clear indications of gas interaction between two velocity components, suggesting an ongoing collision at an estimated angle of $\sim$45$^\circ$ to the line of sight. The estimated collision timescale is 0.35–1.03 Myr, aligned with the inferred ages of young stellar objects (YSOs) in the region, supporting the hypothesis of collision-induced star formation. Hub-filament systems (HFSs) are identified in the compressed gas region, where filaments converge towards a dense hub, suggesting the CCC as a potential driver of HFS and massive star formation. The high column density ($\sim$2$\times 10^{23}$ cm$^{-2}$) suggests favourable conditions for the formation of massive stars. Although alternative kinematic drivers such as longitudinal collapse and shear motion are considered, CCC remains the most plausible explanation for the observed features. Our findings contribute to our understanding of the mechanisms of cloud dynamics and massive star formation in turbulent molecular environments.
}

\maketitle

\section{Introduction}
\label{sect:Introduction}

Star formation is a fundamental process that significantly influences galaxy evolution, primarily occurring within dense molecular gas. Recent theories suggest that this process arises from gravitational instabilities in turbulent molecular clouds, leading to their collapse and the formation of stars \citep{2007ARA&A..45..565M,2007ARA&A..45..481Z}. A high density state is essential for the collapse of molecular clouds and can be achieved through multiple mechanisms that trigger compression. These mechanisms include filamentary mass transport, stellar feedback from supernovae, ionising radiation from massive stars, and dynamic feedback from cloud-cloud collisions (CCCs)  \citep{2013A&A...555A.112P,2017IAUS..316....9M,1978prpl.conf..368H,1994MNRAS.268..291W}. The challenge lies in achieving the ultra-high-density initial conditions required for massive star formation while balancing strong feedback effects, such as radiation pressure exceeding gravitational binding energy. Among the proposed mechanisms, the CCC process effectively addresses the ultra-high-density conditions needed for massive star formation \citep{2018PASJ...70S..58T}.

Numerical simulations demonstrate that CCCs are common in gas-rich galaxies, highlighting their critical role in initiating star formation \citep{2009ApJ...700..358T,2015MNRAS.446.3608D}. Hydrodynamic simulations reveal that shock-compression zones in colliding clouds create dense cores, which act as the initial conditions for star formation \citep{1992PASJ...44..203H, 2010MNRAS.405.1431A, 2014ApJ...792...63T}. 
Supersonic collisions enhance turbulence and amplify magnetic fields, promoting large-scale filamentary structures \citep[e.g.][]{2013ApJ...774L..31I,2018PASJ...70S..53I}. Ultimately, these structures can fragment into dense, massive cores that promote the formation of high-mass stars.

Increasing observational evidence supports the crucial role of CCCs in the formation of diverse stellar structures. CCCs can trigger the formation of stars and clusters \citep[e.g.][]{2020MNRAS.499.3620I,2021PASJ...73S...1F}, including super star clusters \citep{2016ApJ...820...26F}, low- and intermediate-mass stars \citep{2017ApJ...835L..14G}, and massive star clusters \citep{2018PASJ...70S..43S}. Furthermore, O-type stars and their surrounding H{\sc ii} regions formed by CCCs have been observed \citep{2015ApJ...806....7T,2017ApJ...835..142T}. Theoretical and observational comparisons by \cite{2021PASJ...73S...1F} show alignment in the upper mass spectrum of dense cores, summarising findings from over 50 CCC candidates, covering a wide range of stellar masses from individual O stars to massive star clusters ($\sim$10$^{6}$\,\msun) discovered through extensive CO surveys. Thus, recent research increasingly views CCCs as a significant mechanism for high-mass star formation.

\begin{figure*}[h!]
\vspace*{0.2mm}
\centering
\includegraphics[width=0.9\textwidth]{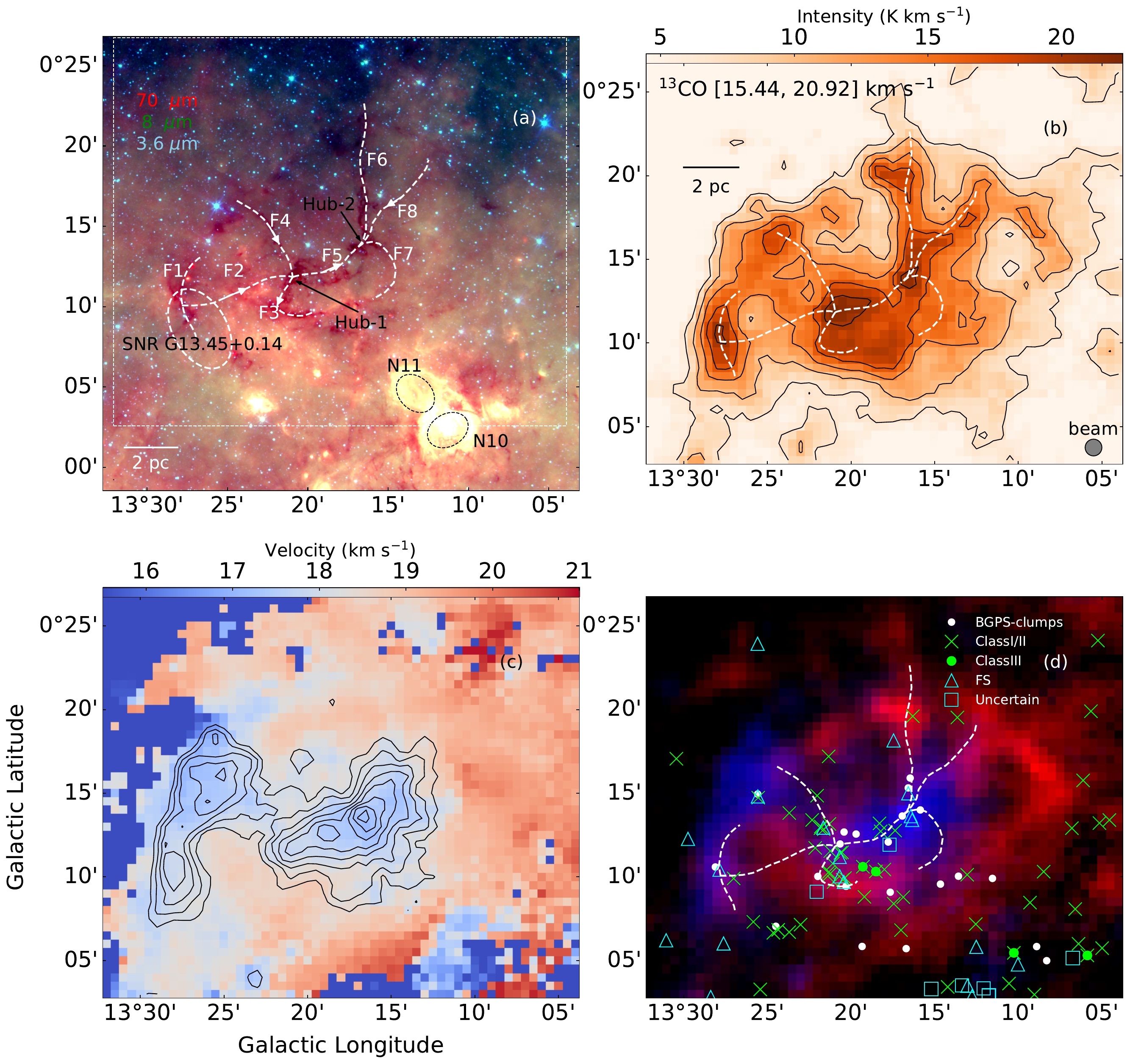}
\caption{(a) Three-colour composite image of G013.313. Blue, green, and red images show the Spitzer IRAC 3.6, 8 \citep{2004ApJS..154....1W}, and Hi-GAL 70\,$\mu$m \citep{2010A&A...518L...1P} data, respectively. The white box with dashed lines shows the target region of the present work.  (b) A map of integrated \xco~\jxs covering the velocity range from 15.44 to 20.92\,\kms\,is shown for the area within the white box in panel (a). Contours correspond to 6.7, 10.0 13.4, 16.7, and 20.1\,K\,\kms. (c) \xco, velocity map (moment-1) of the G013.313 region. Contours correspond to a velocity range of 15.44 to 17.93\,\kms. The contour levels are set at 4.4, 6.0, 7.6, 9.2, 10.8, 12.5, 14.1, and 15.7\,K\,\kms. (d) The two-colour composite image of G013.313 employs \xco~data, with blue indicating the blue-shifted cloud and red representing the red-shifted cloud. The white circles mark the BGPS clumps. Green signs indicate the YSOs identified by \cite{2021ApJS..254...33K}. Crosses represent Class I/II YSOs, while circles denote Class III YSOs. Triangles and squares indicate sources with flat spectral energy distributions and uncertain YSOs, respectively. Dashed white lines display the filament paths identified in Appendix.~\ref{finfinder}.}
\label{G13}
\end{figure*}

CCCs are expected to exhibit several characteristic features, as indicated by theoretical studies and simulations \citep[e.g.][]{1992PASJ...44..203H,2010MNRAS.405.1431A,2014ApJ...792...63T,2018PASJ...70S..58T,2018PASJ...70S..44F,2018ApJ...859..166F,2018PASJ...70S..48H,2018PASJ...70S..43S}. When two clouds of unequal sizes collide, the smaller cloud often forms a cavity within the larger cloud. This interaction results in a complementary spatial distribution of the two clouds at varying velocities. Numerical simulations \citep[e.g.][]{2014ApJ...792...63T,2018ApJ...859..166F} indicate that the collision produces a compressed layer in the interaction region \citep{2021PASJ...73S...1F}, forming a U-shaped cavity orientated towards the smaller cloud. This pattern, as shown in \cite{2018ApJ...859..166F}, may exhibit a displacement between the cavity and the smaller cloud due to projection effects. \cite{2018ApJ...859..166F} present simulations of a small spherical cloud colliding with a larger spherical cloud. If the viewing angle between the line of sight and the collision direction is 0$^\circ$, the red-shifted molecular cloud (larger spherical cloud) may appear as a relatively circular structure in the velocity distribution. At a 45$^\circ$ viewing angle, the red-shifted cloud tends to appear more flattened or elliptical on the velocity map. Furthermore, keyhole-shaped features, characterised by intensity, are frequently observed at CCC sites \citep[e.g.][]{2017ApJ...835..142T, 2018ApJ...859..166F, 2019ApJ...878...26D}. These features, predicted by numerical models, align with synthetic observations \citep{2018ApJ...859..166F}, and their existence has been confirmed through extensive surveys \citep[e.g.][]{2018PASJ...70S..44F,2018ApJ...859..166F,2018PASJ...70S..48H,2018PASJ...70S..43S,2021PASJ...73S.256E,2021PASJ...73S.273F,2024AJ....167..158I,2024PASJ...76..895Y}.

Furthermore, recent observational studies indicate that the early stages of massive star formation often occur at the intersections of dust and molecular filaments, commonly referred to as the Hub-filament system \citep[HFS,][]{2009ApJ...700.1609M}. These structures are believed to play a crucial role in facilitating mass accumulation \citep[e.g.][]{2018ARA&A..56...41M,2019A&A...629A..81T}. The shock-compressed regions resulting from cloud interactions provide favourable conditions for increased gas density and turbulence, which can further lead to the formation of HFSs \citep{2015MNRAS.453.2471B,2018PASJ...70S..53I}. \cite{2024ApJ...974..229M} suggest that the development of HFS in CCC regions arises from a combination of factors, including turbulence, shock compression, magnetic fields, and gravitational effects. Observational evidence for this connection has been reported in both Galactic and extragalactic massive star-forming regions, including SDC13, W31, and W33 complexes, AFGL 5180 and 6366S \citep{2014A&A...561A..83P,2022ApJ...931..115W,2022ApJ...934....2M,2023MNRAS.519.2391Z,2023MNRAS.523.5388M}, as well as N159E-Papillon and N159W-South in the Large Magellanic Cloud \citep{2019ApJ...886...14F,2019ApJ...886...15T}. The diverse structural and kinematic properties of molecular clouds add complexity in assessing the relationship between CCC and HFS formation. Although observational studies have provided insights into this connection, comprehensive analyses remain limited. Expanding the number of observations and systematic analyses is essential for refining our understanding of the physical processes involved and for further clarifying the role of CCC in the development of massive star-forming regions.

This study focuses on G013.313$+$0.193 (G013.313), located in the Galactic plane at \(l=13.313^\circ \) and \(b=0.193^\circ \). This complex region has been observed in various infrared and sub-millimetre surveys, yet studies focusing on its kinematics and related properties remain unexplored. Although previous studies have targeted G013.313, no evidence of massive star formation, such as H{\sc ii} regions or masers, has been documented. Recent observations identified several Spitzer dark clouds within this region \citep{2009A&A...505..405P}, illustrated in the tricolour infrared image shown in Fig.~\ref{G13}a. The image, derived from Spitzer and Hi-Gal data \citep{2004ApJS..154....1W,2010A&A...518L...1P}, displays emissions at 3.6, 8, and 70\,$\mu$m in blue, green, and red, respectively. The region exhibits a rich and complex environment, marked by its proximity to several notable structures. G013.313 lies near the W33 complex at a distance of 2.4\,kpc \citep{2013A&A...553A.117I} and is close in projection to a known filamentary infrared dark cloud located at 3.6\,kpc \citep{2014A&A...561A..83P}. It is also close to the bubbles N10 (4.7\,kpc; \citealt{2006ApJ...649..759C}) and N11 \citep{2015A&A...582A...1D,2020MNRAS.493.2706C}. Furthermore, the supernova remnant (SNR) G13.45$+$0.14, with a velocity of 24\,\kms and a distance of 2.7\,kpc \citep{2021ApJS..253...17S}, is marked by a white ellipse in Fig.~\ref{G13}a. Despite the dynamic surroundings of the region, there is currently no evidence that supports any physical interaction between G013.313 and these nearby structures, including the SNR. 

In our analysis, we used various molecular tracers, including \co, \xco, and \yco~ , all observed in their $J$=1-0 transitions, to investigate the detailed distribution and kinematics of the molecular gas in G013.313. The isotopologues \xco~ and \yco~ are better tracers of relatively dense regions, offering valuable insight into their structure and dynamics. The low-$J$ \hco~lines are sensitive to gas with densities of $\sim$$10^{4}- 10^{6}$ cm$^{-3}$, which are excellent tracers of the velocity field of relatively high-density regions \citep{2003ApJ...597..986Q,2006ApJ...640L.135G}, especially the shock-compressed layer formed due to CCC \citep{2017ApJ...835..142T}. The analysis also includes the ammonia (1,1) and (2,2) lines, which are particularly valuable for revealing the kinetic temperature and density information within molecular clouds, particularly in star-forming environments \citep{1983ARA&A..21..239H, 1983A&A...122..164W, 1988MNRAS.235..229D}. By combining these tracers, we aim to build a more comprehensive understanding of CCCs and their impact on star formation. Such studies will not only contribute to the growing statistics of CCC events, but will also enable a comprehensive picture of the role of CCCs and star formation in the Galaxy.

The paper is structured as follows. Sect.~\ref{sect:Observation} introduces the observational and archival data we employ. In Sect.~\ref{sect:results}, we analyse the spatial and kinematic morphology, as well as the global gas properties, including the identification of clumps and young stellar objects (YSOs) within G013.313. In Sect.~\ref{sec:4}, we further investigate the CCC scenario by analysing its key signatures, including complementary distribution, U-shaped morphology, bridge structures, and V-shaped velocity patterns. We also examine the collision timescale and assess alternative kinematic drivers that could contribute to the observed gas dynamics. Additionally, we explore the role of CCCs in triggering HFSs and evaluate their impact on star and cluster formation in G013.313. Lastly, Sect.~\ref{sect:summary} presents a summary of the main results.

\section{Observations and data reduction}
\label{sect:Observation}

\subsection{CO and \hco observations}
Observations of three CO molecular lines and \hco \,towards G013.313 were carried out between August and November 2023 using the 14 m millimetre-wavelength telescope at the Purple Mountain Observatory in Delingha (PMODLH). 
This antenna can observe three CO molecular lines, \co~\jxs, \xco~\jxs, and \yco~\jxs simultaneously, while \hco~\jxs was observed separately. Observations were undertaken in the on-the-fly (OTF) mapping mode of the position switch. The 3$\times$3 beam Superconducting Spectroscopic Array Receiver (SSAR) system is used as front end, which provides a 1\,GHz bandwidth with 16384 channels and a spectral resolution of 61\,kHz, equivalent to a velocity coverage of $\sim$2600 \kms and a channel width of $\sim$0.17\,\kms at 110\,GHz and 0.19\,\kms at 89\,GHz. The detailed properties of this system are described in \cite{2012ITTST...2..593S}. 

Table~\ref{tab:obs} provides the typical system temperature, beam efficiency ($\eta_{\rm mb}$), half power beam width (HPBW), and channel spacing for the ~\jxs \co, \xco, and \yco~lines. Typical noise temperatures of $\sim$0.54\,K for $^{12}$CO and $\sim$0.26\,K for $^{13}$CO and C$^{18}$O are achieved. Data reduction was performed with CLASS and GREG as part of the GILDAS software package. The final data product was converted to fits file format with a pixel size of 30$\arcsec$ $\times$ 30$\arcsec$. In this work a field of 29$\arcmin$ $\times$ 25$\arcmin$ centred at (l,b)=\(13^\circ\!\!.301, 0^\circ\!\!.244\) is analysed.

\subsection{NH$_3$ observations}
\label{nh3-obser}

We conducted OTF mode observations of NH$_3$ ($J,K$)=(1,1), and (2,2) lines from March 2023 to April 2024 using the Nanshan 26 m telescope. A\,22.0-24.2\,GHz dual polarisation channel superheterodyne was used, with the rest frequency centred on 23.708\,GHz to simultaneously observe NH$_3$\,(1,1) (23.694\,GHz) and (2,2) (23.723\,GHz). 

To convert antenna temperatures $T^\ast_{\rm A}$ into main beam brightness temperatures $T_{\rm MB}$, a beam efficiency of $\sim$0.59 has been applied (see Table~\ref{tab:obs}). The $T_{\rm MB}$ scale uncertainty is about 14\% \citep{2018A&A...616A.111W}. The telescope is equipped with a dual-input digital filter bank (DFB) system with 8192 channels. The typical system temperature is $\sim$50\,K ($T^\ast_{A}$ scale). The Full Width at Half Maximum (FWHM) beam of the telescope is 115$\arcsec$ obtained from point-like continuum calibrators, and the bandwidth is 64\,MHz, resulting in a channel spacing of 0.11\,km\,s$^{-1}$ (listed in Table~\ref{tab:obs}). All velocities are with respect to the Local Standard of Rest (LSR). The total integration time for each on and off position was 360\,s, and some sources required multiple observations to increase the signal-to-noise ratio (S/N). All observations were obtained under good weather conditions and above an elevation of $30\degr$, resulting in a typical RMS noise level of $\sim$20\,mK.

\begin{figure}[h!]
\vspace*{0.2mm}
\centering
\includegraphics[width=0.4\textwidth]
{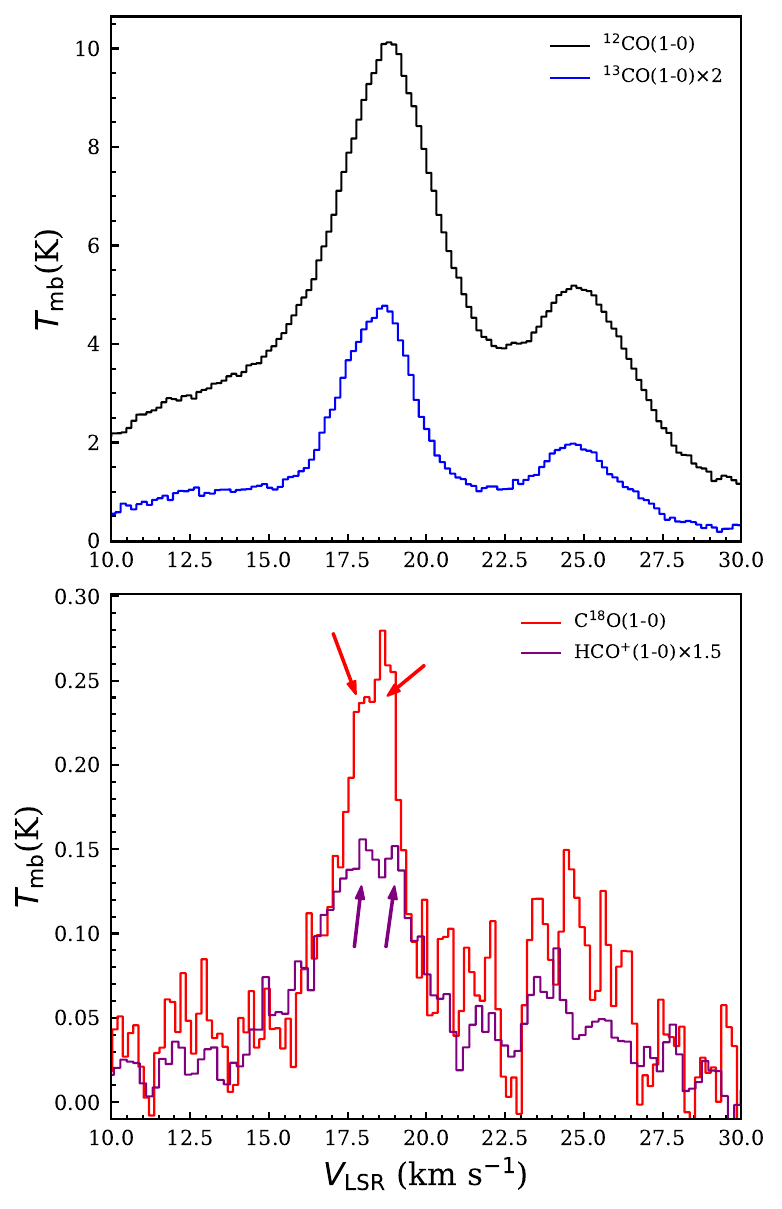}
\caption[]{Spectra for the entire mapped area of the G013.313 region (see Fig.~\ref{G13}b). The upper panel shows the average spectra of \co~(black) and \xco~(blue), while the lower panel displays \yco~(red) and \hco~(purple). In the lower panel, arrows indicate the approximate positions of the two peaks observed in the \yco~ and \hco~ lines (see also Fig.~\ref{G13}c).
}
\label{exempleline}
\end{figure}

\subsection{Data reduction}
The data reduction was performed using the CLASS package of GILDAS \footnote{http://www.iram.fr/IRAMFR/GILDAS/}, the python packages \texttt{matplotlib}  \citep{2007CSE.....9...90H}. A spectrum was considered genuine detection when the S/N was greater than 3. The \NH~data were spectrally smoothed to better compare and analyse them together with the CO data, resulting in a velocity resolution of 0.17\,km\,s$^{-1}$. 
To convert hyperfine blended line widths to intrinsic line widths in the \NH~ inversion spectrum \citep[e.g.][]{1998ApJ...504..207B}, we also fitted the averaged spectra using the GILDAS built-in '\NH\,(1,1)' fitting method, which can fit all 18 hyperfine components simultaneously. From this \NH\,(1,1) fit we can obtain integrated intensity, line centre velocity, intrinsic line widths of individual hyperfine structure (hfs) components, and optical depth.

The main beam brightness temperatures $T_{\rm MB}$ are obtained from the GAUSS fit. Due to the generally weak hyperfine satellite lines of the \NH\,(2,2) transition, the optical depths for \NH\,(2,2) could not be determined. A single Gaussian profile was fitted to the main group of \NH\,(2,2) hyperfine components. Physical parameters of the dense gas such as the rotational temperature ($T_{\rm rot}$), kinetic temperature ($T_{\rm kin}$), and NH$_3$ column density ($N_{\rm NH_{3}}$) were derived in Appendix.~\ref{NH3}.

\begin{table}
\caption{Basic observational parameters at the observed frequencies. \label{tab:obs}}
\centering
\tiny
\begin{tabular}
{lccccccccccc}
\hline\hline
 Lines & Rest frequency& $T_{\rm sys}$  &  $\eta_{mb}$ & HPBW  & channel spacing  \\
 & GHz &(K) & &($''$) & (\kms) \\
\hline
 \co(1-0)   &115.271 & 250-300 & 0.52  & 49   & 0.159  \\
 \xco(1-0)  &110.201 & 150-200 & 0.56  & 51   & 0.166  \\
 \yco(1-0)  &109.782 & 150-200 & 0.56  & 52   & 0.166  \\
 \hco(1-0)  &89.188 & 100-150 & 0.60  & 61   & 0.187  \\
 \NH(1,1)   &23.694& $\sim$50 & 0.59  & 115   & 0.11  \\
 \NH(2,2)   &23.723 & $\sim$50 & 0.59  & 115   & 0.11  \\

\hline  
\end{tabular}
\label{table:average}
\end{table}

\subsection{Archival data}

\begin{figure*}[h]
\vspace*{0.2mm}
\centering
\includegraphics[width=0.97\textwidth]
{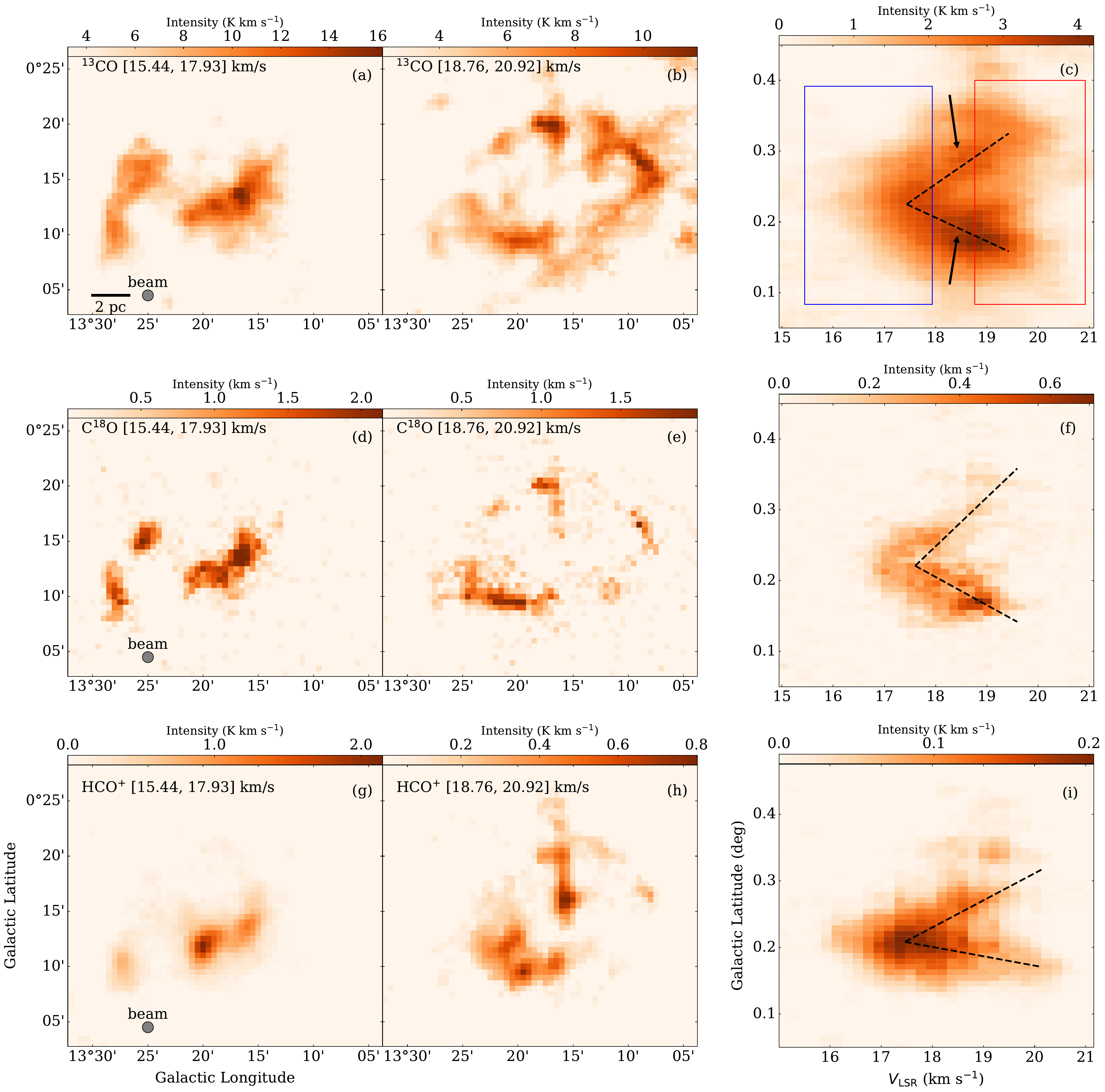}
\caption[]{Integrated intensity maps of the blue and red clouds are displayed in panels (a, b, d, e, g, h), with the velocity range displayed in the upper left corner of each panel. Corresponding P-V diagrams with latitude-velocity data integrated across a longitude range of \(l=13.06^\circ \) to \(l=13.54^\circ \) in panels (c, f, i). The upper three panels (a–c) illustrate the \xco~emission line, while the middle three panels (d–f) depict the \yco~emission line, and the lower three panels (g–i) present the \hco~emission line. Each panel includes the integration ranges indicated in the bottom right corner. V-shaped features are showing in the P-V diagrams (c, f, i), and bridging features are highlighted with arrows in panel (c). Additionally, panel (c) shows the approximate velocity ranges of the two components, with the G013.313-blue component marked by a blue box and the G013.313-red component by a red box.}
\label{moment0}
\end{figure*}

\begin{figure*}[h]
\vspace*{0.2mm}
\centering
\includegraphics[width=0.9\textwidth]
{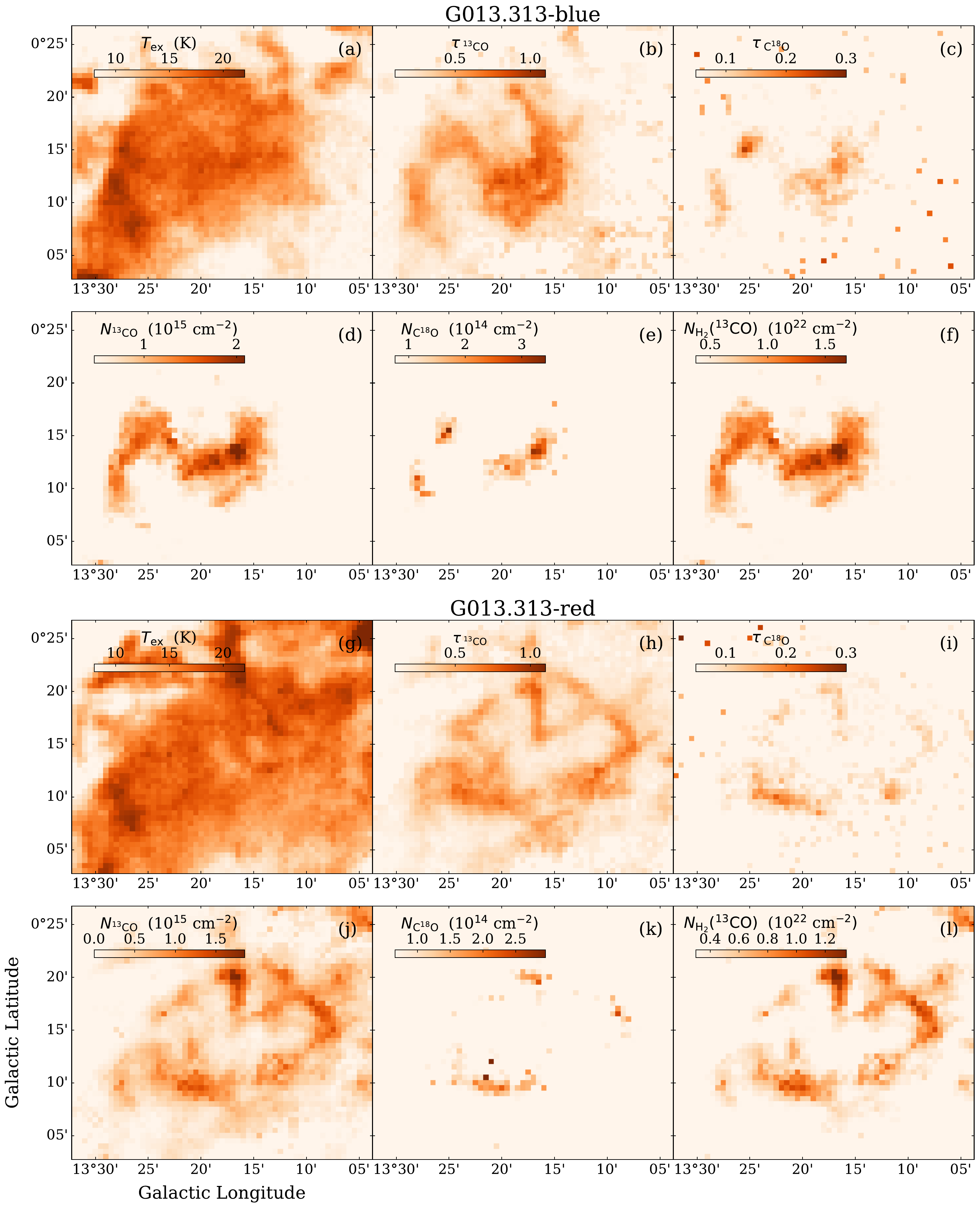}
\caption[]{Distribution of physical parameters in the G013.313-blue region: (a) CO-based excitation temperature map derived from \co. (b) and (c) Optical depth maps of \xco~and \yco. (d) and (e) column Density maps of \xco~and \yco. (f) Hydrogen molecular column density; Distribution of physical parameters in G013.313-red. (g) CO-based excitation temperature map derived from \co. (h) and (i) Optical depth maps of \xco~and \yco. (j) and (k) column Density maps of \xco~and \yco. (l) hydrogen molecular column density.}
\label{calculate}
\end{figure*}

\subsubsection{The infrared data }
Mid-infrared data of the region towards G013.313 were obtained from the archives of the Spitzer Space Telescope \citep{2004ApJS..154....1W}. Images in four IRAC bands centred at 3.6, 4.5, 5.8, and 8.0\,$\mu$m with spatial resolutions ranging from 1.7$\arcsec$ to 2$\arcsec$ were taken from the Galactic Legacy Infrared Midplane Survey Extraordinaire survey (GLIMPSE; \citealt{2003PASP..115..953B}). The 24 and 70\,$\mu$m images with resolutions of 6$\arcsec$ and 18$\arcsec$ were obtained from the MIPS Galactic Plane Survey (MIPSGAL; \citealt{2004ApJS..154...25R, 2009PASP..121...76C}). These data were used to investigate the structure of the region.

Far-infrared data were obtained from the ESA Herschel Space Observatory Infrared Galactic (Hi-GAL) Plane Survey. The Hi-GAL plane survey is an open-time key project \citep{2010A&A...518L...1P, 2010A&A...518L.100M} that mapped the inner Galactic plane at 70 and 160\,$\mu$m with the Photoconductor Array Camera and Spectrometer (PACS; \citealp{2010A&A...518L...1P}) and at 250, 350, and 500\,$\mu$m with the Spectral and Photometric Imaging Receiver (SPIRE; \citealp[]{2010A&A...518L...3G} ). The spatial resolution of the images is 6$\arcsec$ , 12$\arcsec$ , 18$\arcsec$ , 24$\arcsec$ , and 35$\arcsec$ for the five wavelength bands \citep{2010A&A...518L.100M}, respectively.

\subsubsection{Catalogues}
The Bolocam Galactic Plane Survey (BGPS) identifies a population of over 2000 massive clump candidates in an 84.4 deg$^{2}$ region of the first quadrant of the Galaxy, and provides their physical parameters \citep{2016ApJ...822...59S}. The APEX Telescope Large Area Survey of the Galaxy (ATLASGAL) aims to map dense clumps in the Milky Way by using the APEX telescope to observe the Galactic plane, identifying dense clumps.

We adopted YSO candidates from the Spitzer IRAC Candidate YSO (SPICY) Catalog for the Inner Galactic Midplane (\citealp{2021ApJS..254...33K}), which was constructed from wide‑field mid‑infrared imaging by the Spitzer Space Telescope and WISE (e.g. the GLIMPSE survey; \citealp{2003PASP..115..953B,2009PASP..121..213C}).

\section{Results}
\label{sect:results}

\subsection{Kinematics}
\label{kinematic} 

The molecular cloud associated with G013.313 exhibits filamentary structures \citep{2016A&A...591A...5L} that correlate with infrared‐extinction features \citep{2009A&A...505..405P}. In this work a distance of 2.1\,kpc, as measured in previous studies \citep[e.g.][]{2006ApJS..163..145J, 2010ApJ...718L.106B, 2011ApJ...731...90D, 2014MNRAS.437.1791U, 2015A&A...579A..91W}, is adopted. Fig.~\ref{G13}b illustrates the integrated intensity distribution of the \xco~\jxs emission over a velocity range of 15.44 to 20.92\,\kms, indicated by the white box in Fig.~\ref{G13}a. The \xco~contours delineate the 8\,$\mu$m infrared dark cloud, offering insights into the spatial distribution of the molecular gas.
The presence of bright spots at 70\,$\mu$m in the three-colour image of G013.313 (Fig.~\ref{G13}a) supports the existence of protostellar objects within this region.

Fig.~\ref{exempleline} shows the averaged spectra of \co~(black), \xco~(blue), \yco~(red), and \hco~(purple) across the entire mapped region (see Fig.~\ref{G13}b). In particular, within the velocity range of G013.313 (10-30\,\kms), both \co~and \xco~exhibit a single emission peak. Line profiles in Fig.~\ref{exempleline} show the superposition of two spectral components, which are indicated by arrows in the lower panel and are disentangled in the moment-1 velocity maps of Fig.~\ref{G13}c. This variation arises because \yco~and \hco~can trace deeper into the dense region of the cloud, capturing more complex velocity structures that remain unresolved by \co~and \xco. These double peaks suggest the presence of two distinct velocity components, likely representing separate cloud structures. 

Notably, both the \co~(black) and \xco~(blue) lines exhibit broad wings, particularly around $\sim$15 \kms, with broad spectral extensions at high and low velocities. Such wing structures are often considered key evidence of CCC \citep{2014ApJ...792...63T}, possibly associated with turbulence enhancement or shock-induced high-velocity gas components. However, since the region also contains YSOs and clumps (see Sects.~\ref{sec:clump} and \ref{YSO}), complex dynamical processes could contribute to spectral profiles \citep{2017ApJ...835L..14G,2019A&A...632A.115G}, making it difficult to attribute the wing features solely to CCC. Additionally, the wings may also be related to the presence of a velocity bridge, as the bridge represents a continuous connection between the two velocity components, which could manifest itself as an extended spectral feature. A detailed discussion is provided in Sect.~\ref{bridge}. Furthermore, as also shown in Fig.~\ref{exempleline}, another velocity component is visible around 25\,\kms. This component lies within the range typically associated with the Sagittarius spiral arm in this region \citep{2016ApJ...823...77R} and corresponds to the SNR G13.45$+$0.14 mentioned earlier in Sect.~\ref{sect:Introduction}.
We created a position-velocity (P-V) diagram (Fig.~\ref{pv} in Appendix.~\ref{appendB}) for the mapped area with a velocity range of 15-28\,\kms. This confirms that our target velocity range (15.44-20.92\,\kms) shows no velocity correlation with the velocity component at $\sim$25\,\kms.

The velocity distribution of G013.313 reveals considerable variation,  (indicated in Figs.~\ref{channel-co}-\ref{channel-hco} of Appendix.~\ref{appendB}, which show channel maps of \co~(background) with \xco~contours overlaid at intervals of 0.31\,\kms, along with channel maps of \hco. From the \xco~ contours of the channel maps, it is evident that the gas between $V_{\rm LSR}$ = 15.40 and 17.94\,\kms shows the presence of an 'S' structure, which gradually appears and then dissipates. Meanwhile, the gas in the velocity range $V_{\rm LSR}$ = 18.25 to 20.48\,\kms reveals a 'U' structure, which opens towards higher Galactic longitudes. The 'S' structure is primarily distributed in the Galactic western region (towards decreasing Galactic longitude), whereas the 'U' structure is concentrated in the Galactic eastern region (towards increasing Galactic longitude).

Fig.~\ref{G13}c shows the velocity field, the first-moment maps of \xco~across the velocity range from 15.44 to 20.92\,\kms. This map reveals the spatial distribution of the two distinct velocity components. In particular, the separation between the blue-shifted (15.44-17.93\,km\,s$^{-1}$) and red-shifted clouds (18.76-20.92\,km\,s$^{-1}$) is evident, highlighting the region where the velocity ranges diverge. The distinct velocity structures can be easily seen, allowing for a clear visualisation of the spatial and kinematic differences between the two clouds.

By integrating spectral, moment and channel map analyses, we determine that the blue-shifted cloud occupies a velocity range from 15.44 to 17.93\,\kms, with a mean velocity of 17.6\,\kms, whereas the red-shifted cloud spans a velocity range of 18.76 to 20.92\,\kms, with a mean velocity of 19.5\,\kms. This results in a velocity separation of $\sim$2\,\kms along the line of sight between the two clouds. However, these clouds are not completely distinct and exhibit some overlap in the velocity space. We designate the two regions as G013.313-blue and G013.313-red, respectively, with a transition region between them (referred to as the 'bridge' in Sect.~\ref{bridge}).

\subsection{Morphology}
\label{morphology} 

Figure~\ref{moment0}a and b illustrate the spatial distribution of \xco~\jxs. The G013.313-blue component mainly exhibits compact emission, occupying an area of approximately 11\,pc~$\times$\,11\,pc in an 'S' shape around G013.313 (Fig.~\ref{moment0}a). The G013.313-red component presents a more diffuse emission pattern, extending over a region of about 12\,pc\,$\times$\,9\,pc in a 'U' shape (Fig.~\ref{moment0}b). Fig.~\ref{moment0} c shows the Galactic latitude versus velocity diagram of G013.313. The velocity distribution shows a V-shape with a maximum velocity of 21\,\kms. The distribution is denser at the pointed tip of the V-shaped structure compared to other regions. The two clouds are connected in the velocity space by 'bridging features' with intermediate velocities (17.93-18.76\,\kms), the locations of which are indicated by arrows in Fig.~\ref{moment0}c. 

Figs.~\ref{moment0}d-e illustrate the spatial distribution of \yco~\jxs in the G013.313 region. The G013.313-blue component (Fig.~\ref{moment0}d) forms isolated patches of emission, lacking the cohesive structure seen in \xco. Its intensity is notably lower compared to the intensity distribution of the G013.313-blue component (Fig.~\ref{moment0}a) . The G013.313-red component (Fig.~\ref{moment0}e) shows a fragmented U-shaped pattern, the emission is not well connected, with clear gaps between different regions. Fig.~\ref{moment0}f presents the Galactic latitude–velocity diagram for \yco, where the velocity distribution also displays a V-shape, although it is not very extended. The denser region is located near the apex of the V-shape, but is less pronounced compared to \xco. 

Figs.~\ref{moment0}g-h illustrate the spatial distribution of \hco~\jxs in the G013.313 region. The G013.313-blue component reveals partial emission that resembles one-half of the 'S' shape seen in the \xco~data, the emission is more localised and less extended. The red component of \hco~shows strong emission concentrated in the central part of the 'U' shape, where CO emissions are more diffuse. This indicates that \hco~traces denser and more confined regions within the molecular gas, particularly in the core areas that are less prominent in CO observations. The P-V diagram of \hco~(Fig.~\ref{moment0}i) shows V-shaped velocity structure similar to CO. There is a noticeable dip at the centre of the V, where the emission is densest.

The velocity separation between G013.313-blue and G013.313-red is as mentioned in Sect.~\ref{kinematic} $\sim$2\,\kms. Based on Equation.~\ref{soundspeed} in the Appendix.~\ref{NH3}, and using the average kinetic temperature $\sim$16\,K (in Sect.~\ref{parameters}), the calculated sound speed for G013.313 is $\sim$0.31\,\kms, confirming that the motion is supersonic. Furthermore, the spatially complementary distribution of the two molecular components and the presence of a velocity bridge along their interface suggest complex dynamical interactions between two components.

\subsection{Physical parameters}
\label{parameters} 

To analyse the physical conditions within the G013.313-blue and G013.313-red regions, we calculated several key parameters such as excitation temperature, optical depth, and column density. The methods for CO-related calculations are provided in Appendix.~\ref{appendA}, while those for ammonia molecular lines are detailed in Appendix.~\ref{NH3}. In the following, we discuss the resulting maps and values.

Figs.~\ref{calculate}a and~\ref{calculate}g illustrate the excitation temperature maps derived from the CO lines for G013.313-blue and G013.313-red, respectively. CO-based excitation temperatures range for G013.313-blue from 2.72 to 22.03\,K, with an average value of 12.38$\pm$1.01\,K, while G013.313-red spans from 2.71 to 25.43\,K, averaging at 14.68$\pm$1.01\,K.

The optical depths of the \xco~and \yco~lines for G013.313-blue are in ranges 0.08 < $\tau\,_{^{13}\mathrm{CO}}$ < 0.88 and 0.05 < $\tau\,_{\mathrm{C}^{18}O}$ < 0.26, respectively. For G013.313-red, the optical depths are slightly different: 0.09<$\tau\,_{^{13}\mathrm{CO}}$ < 0.77 and 0.04 < $\tau\,_{\mathrm{C}^{18}\mathrm{O}}$ < 0.38. Figs.~\ref{calculate}b-c display the optical depth maps for the \xco~and \yco~emissions in G013.313-blue, while Figs.~\ref{calculate}h-i show the corresponding maps for G013.313-red. These maps indicate that the \xco~emission is moderately optically thin in most of the observed region. Furthermore, the optical depth distribution maps of \xco~and \yco~clearly reveal two distinct structures. In particular, the distribution of $\tau\,_{^{13}\mathrm{CO}}$ highlights the similarity in optical depths between the G013.313-blue and G013.313-red clouds, suggesting comparable physical conditions across both regions.

The column density of \(\mathrm{^{13}CO}\) for G013.313-blue spans a range of \(0.19 - 23.19\,\times\,10^{15} \,\mathrm{cm}^{-2}\)(Fig.~\ref{calculate}d), with an average value of \(5.10\pm0.54 \times\,10^{15} \,\mathrm{cm}^{-2}\). In comparison, the column density measured by \(\mathrm{C^{18}O}\) covers a smaller range of \(0.14 - 3.80 \times\,10^{15} \,\mathrm{cm}^{-2}\)(Fig.~\ref{calculate}e), with an average of \(1.02\pm0.10 \times\,10^{15} \,\mathrm{cm}^{-2}\). For G013.313-red, the column density derived from \(\mathrm{^{13}CO}\) is in the range \(0.21 - 18.55 \times\,10^{15}\,\mathrm{cm}^{-2}\)(Fig.~\ref{calculate}j), with an average of \(4.01\pm0.04\times\,10^{15} \,\mathrm{cm}^{-2}\), while the \(\mathrm{C^{18}O}\) column density ranges in \(0.12 - 3.29 \times\,10^{15} \,\mathrm{cm}^{-2}\)(Fig.~\ref{calculate}k), with an average of \(0.70\pm0.07 \times\,10^{15} \,\mathrm{cm}^{-2}\). Additionally, the H$_2$ column density derived from \(\mathrm{^{13}CO}\) for G013.313-blue spans a range of \(0.01 - 18.71 \times 10^{22} \,\mathrm{cm}^{-2}\)(Fig.~\ref{calculate}f), with an average of \(2.06 \times 10^{22} \,\mathrm{cm}^{-2}\). For G013.313-red, the H$_2$ column density ranges from \(0.01 - 14.97 \times 10^{22} \,\mathrm{cm}^{-2}\)(Fig.~\ref{calculate}i), with an average of \(2.39\pm0.35 \times 10^{22} \,\mathrm{cm}^{-2}\). The masses of G013.313-blue and G013.313-red, are approximately 0.86 and 1.10\,$\times$\,10$^{5}$\,\msun, respectively. 

\begin{figure*}[h!]
\vspace*{0.2mm}
\centering
\includegraphics[width=0.9\textwidth]
{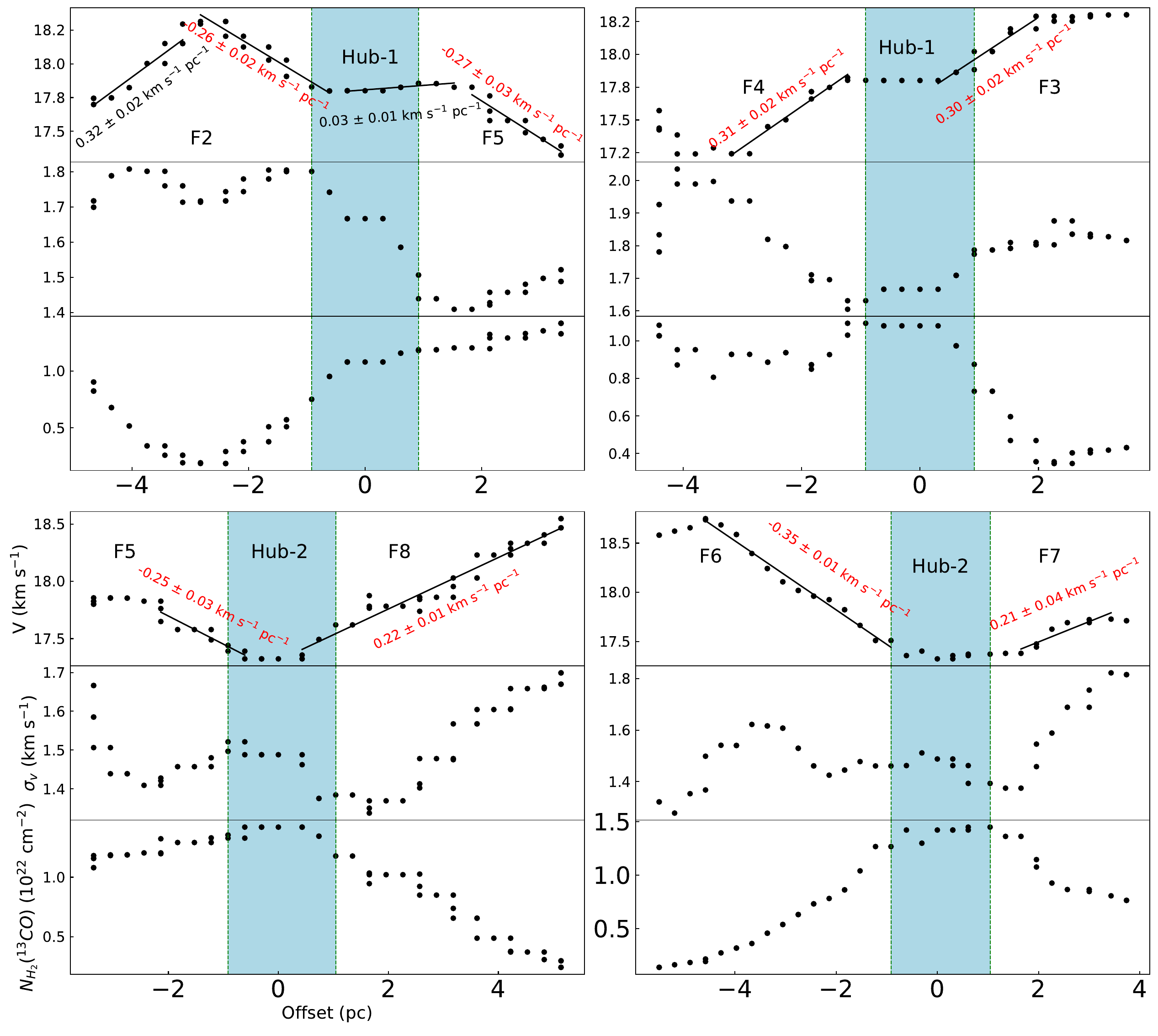}
\caption[]{Distributions of velocity, velocity dispersion, and column density variations along the filament skeletons marked in Fig.~\ref{G13}. The top-left panel shows the profiles along F2–F5, while the top-right panel displays F4–F3. The middle-left panel represents F5–F8, and the middle-right panel corresponds to F6–F7. The direction of each profile follows the arrows shown in the Fig.~\ref{G13}a. The figure also highlights the approximate regions of Hub-1 and Hub-2, derived from the intensity distribution in Fig.~\ref{G13}b.}  
\label{inflow}
\end{figure*}

The kinetic temperatures and non-thermal velocity dispersions of the two clouds are similar, with kinetic temperatures of 15.81$\pm$5.8\,K and 16.01$\pm$4.3\,K, and velocity dispersions of 0.89$\pm$0.02\,\kms and 0.81$\pm$0.01\,\kms, respectively.  Additionally, the relatively high Mach numbers for both molecular clouds indicate that, despite the modest non-thermal velocity dispersions, the dynamics in these regions are indeed non-thermally dominated. This highlights the complex interplay of thermal and non-thermal processes in the molecular gas dynamics of these clouds. These values, along with the average values of other relevant parameters for the two molecular clouds, are summarised in Table~\ref{tab:2parameter}. 

\begin{table}
\caption{Physical parameters of the molecular clouds. 
\label{tab:2parameter}}
\centering
\begin{tabular}
{lcccccccccccc}
\hline\hline
Parameter & G013.313-blue & G013.313-red \\
\hline
$V_{\rm LSR}$ & 15.45-17.93 & 18.76-20.92 \\
$V_{\rm center}$ & 17.6 & 19.5 \\
Size & 11\,pc\,$\times$\,11\,pc  &  12\,pc\,$\times$\,9\,pc\\
$\Delta v$ & 0.97$\pm$0.17 & 0.87$\pm$0.25 \\
$T_{\rm ex}$ & 12.38$\pm$1.01 & 14.68$\pm$1.01 \\
$N_{\rm ^{13}CO}$(peak), 10$^{16}$\,\cmm & 2.32 & 1.86 \\
$N_{\rm C^{18}O}$(peak), 10$^{15}$\,\cmm & 3.80 & 3.29 \\
$N_{\rm H_{2}}$$(\rm ^{13}CO$,peak), 10$^{23}$\,\cmm & 1.87 & 1.50\\
$M_{\rm iso}$, 10$^{5}$\,\msun & 0.86$\pm$0.09 & 1.10$\pm$0.08 \\
$T_{\rm kin}$ & 15.81$\pm$5.80 & 16.01$\pm$4.25 \\
$N_{\rm NH_{3}}$(peak), 10$^{15}$\,\cmm  & 2.31$\pm$0.25 & 1.48$\pm$0.16\\
$\sigma_{\rm NT}$ & 0.89$\pm$0.02 & 0.81$\pm$0.01 \\
$\mathcal{M}$ & 3.92$\pm$0.18 & 3.45$\pm$0.12 \\
\hline  
\end{tabular}
\tablefoot{Row 1: Names of clouds. Row 2: velocity ranges. Row 3: Peak velocities. Row 4: describes the extent of the respective cloud. Row 5: Velocity line widths derived from the 2nd moment map. Row 6: Mean values of CO-based excitation temperatures for each cloud. Row 7: Maximum \xco~column density towards each cloud. Row 8: Maximum \yco~column density towards each cloud. Row 9: Maximum hydrogen column density derived from \xco~towards each cloud. Row 10: Total molecular mass of two clouds. Row 11: Mean values of kinetic temperature derived from \NH. Row 12: Maximum \NH~column density towards each cloud. Row 13: Mean values of non-thermal velocity dispersion derived from \NH. Row 14: Mean values of mach number for each cloud derived from \NH.}
\end{table}

Excitation temperatures, optical depths, kinetic temperatures, and non-thermal velocity dispersions indicate that the G013.313-blue and G013.313-red clouds exhibit similar general physical conditions. However, the \xco~column density maps in Figs.~\ref{calculate}d and \ref{calculate}j reveal that G013.313-blue exhibits higher column densities in its inner region, while G013.313-red shows comparatively lower values. There is a noticeable difference in ammonia column densities between the two clouds. The blue-shifted molecular cloud shows a higher ammonia column density of 2.31$\times$10$^{15}$\cmm, compared to the red-shifted cloud, which has a column density of 1.48$\times$10$^{15}$\cmm. This difference suggests that the blue-shifted cloud has a higher proportion of mass concentrated in its central dense regions, which is further supported by
evidence of gas compression due to CCCs, as discussed in Sect.~\ref{sec:CCC}. This contrast is also likely due to the presence of an HFS in G013.313-blue (Sect.~\ref{sec:hub}), where multiple filament branches converge, facilitating material accumulation. In contrast, G013.313-red lacks such a pronounced convergent structure, leading to lower overall column densities. Additionally, the H$_2$ column density distributions in Figs.~\ref{calculate}f and \ref{calculate}i further support this interpretation. The implications of these density structures, along with their kinematic signatures, are explored in detail in Sect.~\ref{HFS-CCC}.

\subsection{Hub–filament systems}
\label{sec:hub}
A detailed examination of the Spitzer 8 $\mu$m extinction features (Fig.~\ref{G13}b) and the molecular gas distribution reveals the presence of filamentary structures within the target region. To quantitatively identify these filaments, we applied the FilFinder algorithm to Spitzer IRAC 8$\mu$m emission and \xco~(1-0) data within the velocity range of 17.93–18.76\,\kms (see Appendix.~\ref{finfinder}). The resulting filament skeletons are presented in Fig.~\ref{fig:filfinder} and are highlighted in Fig.~\ref{G13}a,b,d. The analysis identifies multiple filaments on the parsec scale that converge at two distinct junctions, consistent with the definition of HFSs as described by \cite{2009ApJ...700.1609M}. Consequently, we classify two HFSs in G013.313, designated as HFS-1 and HFS-2. The central hub regions associated with these systems are labelled Hub-1 and Hub-2 in Fig.~\ref{G13}a. From left to right, we identify eight filaments within the region, labelled F1 to F8. Among them, the filaments F2, F3, F4, and F5 converge towards Hub‑1, while F5, F6, F7, and F8 converge towards Hub‑2. In particular, F5 serves as a bridge, physically connecting the two hub regions.

It is important to emphasise that, while we identify eight probable filaments within the cloud on the basis of our data, their identification is influenced not only by the resolution of the observations but also by velocity projection effects. The observed filamentary structures may appear more coherent or fragmented depending on the line-of-sight velocity components, potentially leading to an underestimation of their complexity. Future high-resolution studies, combined with velocity-resolved analyses, could reveal additional substructures or intricate gas flows within the filaments \citep[e.g.][]{2021ApJ...908...70H}. However, we proceed with the current data to characterise these structures and assess their role in cluster formation within the cloud.

\begin{figure}[h!]
\vspace*{0.2mm}
\centering
\includegraphics[width=0.45\textwidth]
{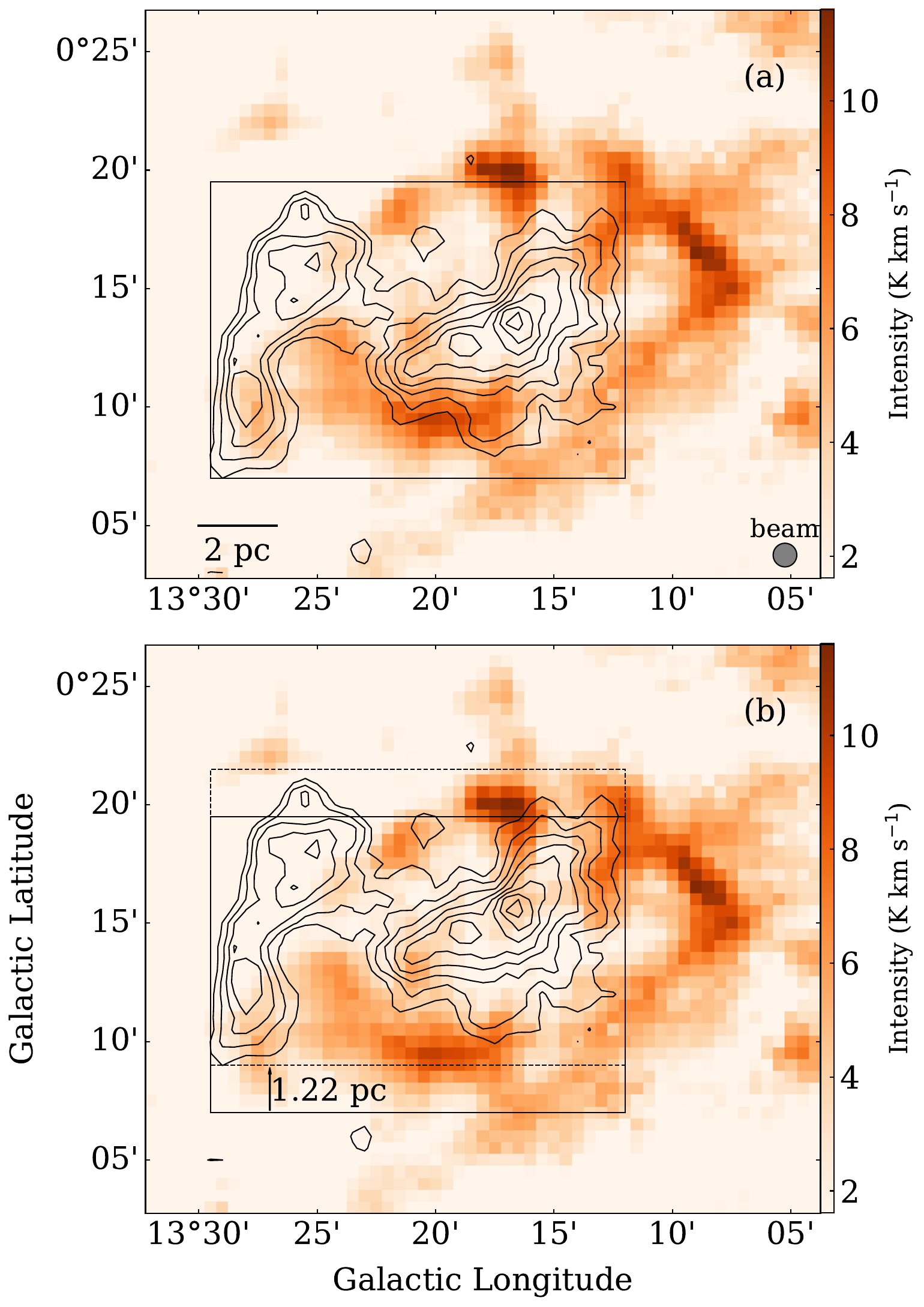}
\caption[]{Complementary \xco~distributions of the two velocity components of G013.313. The colour image represents the integrated velocity range of 18.76 to 20.92\,\kms. Contours correspond to a range of 15.5 to 17.9\,\kms. Contour levels are set at 4.4, 6.0, 7.6, 9.2, 10.8, 12.5, 14.1, and 15.7\,K\,\kms. The colour image highlights the red-shifted cloud, whereas the contours depict the blue-shifted cloud. Panels (a) and (b) show the maps before and after the displacement of the blue-shifted cloud contours, respectively.}
\label{displace}
\end{figure}

\subsection{Filament velocity gradient}\label{gradient}

In Fig.~\ref{G13}c, the moment-1 map reveals a clear velocity gradient across G013.313, extending from G013.313-blue to G013.313-red. This global velocity trend provides an essential foundation for analysing velocity gradients along individual filamentary structures within the region. By examining the velocity variations along the filament skeletons, we aim to further investigate the kinematic properties and potential gas flows that shape the HFS \citep{2018ARA&A..56...41M}.

We extracted data from the filamentary structures connected to the two hubs, Hub-1 and Hub-2. In Fig.~\ref{inflow}, we present the velocity, velocity dispersion, and column density variations along these filaments. For Hub-1, both F2–F5 and F4–F3 exhibit smooth velocity gradients with slight variations at the boundary of the hub, ranging between 0.03 and 0.32 km s$^{-1}$ pc$^{-1}$. Hub-2 shows more pronounced velocity variations, with gradients ranging from 0.21 to 0.35 km s$^{-1}$ pc$^{-1}$, forming a clear V-shaped structure indicative of converging gas flows \citep{2019A&A...629A..81T}. These gradients are consistent with those detected in several other HFS, such as 0.15–0.6 km s$^{-1}$ pc$^{-1}$ in SDC13 \citep{2014A&A...561A..83P}, and 0.14-0.44 km s$^{-1}$ pc$^{-1}$ in G326.611+0.811  \citep{2023ApJ...957...61H}.

Regarding the velocity dispersion, Hub-1 shows a gradual decrease along F2–F5 from 1.8 to 1.4 km s$^{-1}$, while along F4–F3, the dispersion decreases within the hub region. For Hub-2, the velocity dispersion does not exhibit abrupt fluctuations, although moderate variations are persistent within the hub region.

Column density variations are evident in both hubs, with peak values occurring near the centres of the hubs. This is consistent with the material accumulation scenarios \citep{2010A&A...520A..49S}. Hub-2, in particular, displays a increase in column density, supporting the presence of active material accumulation and indicating its crucial role in mass accretion and star formation processes.

\subsection{Dense clumps}
\label{sec:clump}
We have identified 21 BGPS dust continuum clumps at 1.1\,mm (from \citealt{2016ApJ...822...59S}) and 10 ATLASGAL dust continuum clumps at 870\,$\mu$m (from \citealt{2018MNRAS.473.1059U}) towards the G013.313 region, with all 10 clumps from these ATLASGAL clumps overlapping the BGPS clumps (see Fig.~\ref{G13}d). To characterise the properties and dynamics of these clumps, we employed \xco~data, which serve as a more reliable tracer of denser gas and have been found to be optically thin in G013.313 (see Sect.~\ref{parameters}). We plot the spectra extracted around each clump within a given radius (Fig.~\ref{spec-clumps} of Appendix.~\ref{append-clump}). In particular, the spectra of two clumps, B2831 and B2830, show noise rather than distinct emission lines, indicating that they may be located in different spatial regions not directly associated with the molecular cloud studied. Some clumps contain multiple velocity components, which complicates our analysis.

Line asymmetry in molecular spectra serves as a vital diagnostic of gas dynamics. We classify the different non-Gaussian line profiles of our clumps using the following designations: BP (blue profile), RP (red profile), BA (blue asymmetric), RA (red asymmetric), BW (blue wing), RW (red wing), and P (pedestal). BP (blue profile) and RP (red profile) refer to double-peaked profiles with a central absorption dip. In blue profiles, the blue-shifted peak is stronger than the red one, typically indicating infall motions resulting from foreground self-absorption \citep{1993ApJ...404..232Z,1994ASPC...65..192M, 1996ApJ...465L.133M, 1997ApJ...489..719M}. In contrast, red profiles exhibit a stronger red-shifted peak, usually associated with expansion or outflow \citep{1997ApJ...489..719M}. BA (blue asymmetric) profiles are characterised by a single peak near the systemic velocity with a pronounced extension on the blue side, suggesting slight infall or complex velocity components. RA (red asymmetric) profiles are similar but show redward asymmetry, often linked to expansion or rotational motion. \citep{2019A&A...626A..84E, 2005A&A...439.1023C}. BW (blue wing) and RW (red wing) profiles show high-velocity emission tails that extend beyond the main peak on the blue and red sides and are commonly interpreted as indicators of molecular outflows \citep{1999ASIC..540..227B}. P (pedestal) profiles exhibit a broad, low-intensity base underlying the main spectral component, typically spanning several km/s, and are thought to result from large-scale turbulence, weak shocks, or overlapping velocity structures \citep{2012A&A...542A..86Y}.

Most of the clumps exhibit complex asymmetric profiles.  Among the 21 BGPS clumps, three clumps exhibit blue asymmetric profiles (B2865, B2858 and B2869); one exhibits a red asymmetric profile (B2897); two show red profiles (B2862 and B2868); six display blue wing profiles (B2861, B2880, B2852, B2886, B2848 and B2900); two have blue profiles (B2877 and B2881); and one clump presents a pedestal profile (B2882). The remaining six clumps, B2841, B2876, B2864, B2893, B2831, and B2830, are marked with an asterisk, indicating that their spectral profiles are ambiguous and cannot be confidently classified. These clumps are located outside the typical G013.313-blue or G013.313-red regions, and their spectra display complex features that hinder a straightforward interpretation of their dynamic state. The detailed profile classification for each clump is provided in Table~\ref{clumps} in Appendix.~\ref{append-clump}.

We also provide the parameters of the 21 BGPS clumps analysed in this study (Table~\ref{clumps} of Appendix.~\ref{append-clump}). The names, coordinates, and sizes of the clumps are taken from the BGPS. The parameters of the molecular spectral line are derived from the \xco~data, based on the central positions and sizes of the clumps. The table includes key parameters such as central velocity, line width, \xco~column density, molecular hydrogen column density, and mass. The calculations of \xco~column density, hydrogen molecular column density, gas mass, and virial mass are the same as in Appendix.~\ref{appendA}.

The masses of the 21 BGPS clumps in Appendix.~\ref{append-clump} range from 7 to 656\,\msun, while their virial masses vary from 168 to 2169\,\msun. Line-of-sight velocities ($V_{\rm LSR,^{13}CO,clump}$) of the clumps range from 17.17 to 19.59\,\kms. Line widths ($\Delta v_{\rm ^{13}CO,clump}$) range from 0.87 to 1.79\,\kms, while the clump radii ($R_{\rm clump}$) range from 0.1 to 0.7\,pc.

\subsection{Young stellar population}
\label{YSO} 

Using the SPICY catalogue, we identified 94 YSOs within the G013.313 region. Among these, 64 YSOs were classified as Class I and Class II, indicating early stages of stellar evolution, while 4 YSOs were classified as class III, representing more evolved objects. Additionally, we identified 7 uncertain YSOs and 19 sources with spectral energy distributions (SEDs), where the emission spectrum appears roughly flat across the 4.5--24\,$\mu$m. YSOs are marked in Fig.~\ref{G13}d, where class I and class II YSOs are represented by green crosses and circles, flat SED sources are marked by light blue triangles, and the 7 uncertain YSOs are marked by light blue squares. Most YSOs are located along the compressed interface between G013.313-blue and G013.313-red, as well as within the hubs and along the filaments. Notably, Hub-1 hosts a larger number of YSO candidates compared to Hub-2.

\section{Discussion}\label{sec:4}

\subsection{Signatures of CCC}
\label{sec:CCC} 
Several key observational signatures of CCCs have been proposed—complementary spatial distributions with displacement, 'bridge' structures (which can sometimes exhibit a V shape), and U‑shaped gas features \citep{2021PASJ...73S...1F}. We apply these diagnostics to the G013.313 region to evaluate the role of CCCs in shaping its molecular environment.

\subsubsection{Complementary distribution and U shape}
\label{complementary}
Our analysis of the G013.313 region reveals a complementary distribution and distinct U-shaped morphology, indicative of a CCC event. In Sect.~\ref{sect:results}, we present the kinematic, morphology, and physical parameters of the G013.313 region. From these results, the presence of a U-shaped structure is evident and a keyhole configuration can be readily identified. Specifically, G013.313-blue corresponds to the 'key' structure, while G013.313-red aligns with the 'hole' structure. This observation is consistent with CCC scenarios, where denser gas regions undergo deformation due to impact, forming the characteristic U shape, a key observational feature of CCCs as suggested by \cite{2021PASJ...73S...1F}. Furthermore, at the intersection of the 'U'-shaped and 'S'-shaped structures, a compressed region is visible, particularly in the velocity range of 17.94–18.57\,\kms\,in the channel maps of \xco~and \hco~(Figs.~\ref{channel-co}-\ref{channel-hco}). In this region, we observe not only strong emission but also changes in optical depth, density, and other physical parameters along this specific direction. These structural features collectively provide strong evidence of CCC in the G013.313 region, highlighting the dynamic interaction between clouds in this environment. The enhanced emission and physical conditions in these regions strongly suggest that the observed features are a direct result of such interactions.

Our observations of the velocity distribution and the P-V diagrams strongly suggest a viewing angle of approximately 45$^\circ$ for the CCC. In our velocity distribution map (Fig.~\ref{G13}c), the red-shifted cloud exhibits an elliptical shape (U-shaped), consistent with the theoretical predictions described by \cite{2018ApJ...859..166F} for a collision geometry observed at approximately 45$^\circ$. This interpretation is reinforced by the P-V diagrams of dense molecular clouds, the V-shaped structure at a 45$^\circ$ viewing angle is often not symmetric but rather skewed towards the red-shifted cloud. This is also evident in Fig.~\ref{moment0} (panels c, f, and i), where the V-shaped vertex clearly shifts towards the red-shifted cloud (G013.313-red), with its vertex located to the right edge of the blue box corresponding to the G013.313-blue cloud in Fig.~\ref{moment0}c. These observations suggest that our viewing angle is likely not 0$^\circ$, but somewhere between 0$^\circ$ and 90$^\circ$, with a preference towards 45$^\circ$.

\subsubsection{Bridge and V shape}
\label{bridge}
We identify bridge and V-shaped features in the position–velocity diagrams, which provide compelling evidence for a CCC in G013.313. In the context of a CCC, the P-V diagrams frequently reveal distinctive features such as 'bridge' or V-shaped structures, which result from momentum exchange during the interaction. Demonstrated through synthetic P-V diagrams in CCC models, these features are regarded as clear indicators of CCCs \citep[e.g.][]{2015MNRAS.454.1634H,2015MNRAS.450...10H}. Observational studies have also consistently confirmed the diagnostic value of these features in various CCC systems \citep{2017ApJ...835L..14G,2020MNRAS.499.3620I,2021PASJ...73S.273F,2022A&A...663A..97M}. 

We observe V-shaped structures in the PV diagrams across the $J$=1-0 lines of \xco, \yco, and \hco, as shown in Figs.~\ref{moment0}c, f and i. In Fig.~\ref{moment0}c, the blue and red boxes represent the velocity ranges of the G013.313-blue and G013.313-red molecular clouds, respectively. The arrows indicate a distinct bridge feature connecting the blue- and red-shifted components of the two clouds. This bridge feature is a clear indication of the interaction between G013.313-blue and G013.313-red. The presence of this bridge in the velocity space strongly suggests that these two clouds are undergoing a CCC.

\subsection{Collision timescale}
\label{sec:time} 
We estimate the timescale of the collision event based on the spatial and velocity separation between the two clouds. As determined via visual inspection, the two components show complementary distributions with a displacement in space as shown in Fig.~\ref{displace}, where the contours correspond to a range of 15.44 to 17.93\,\kms\,(G013.313-red), while the colour image represents the integrated velocity range of 18.76 to 20.92\,\kms\,(G013.313-red). Fig.~\ref{displace}a shows an overlay of the two components and Fig.~\ref{displace}b shows the two components where a displacement in Galactic latitude is applied, which aims to correct for the collisional motion tilted to the line of sight \citep{2021PASJ...73S...1F}. The estimated displacement along the Galactic latitude is $\sim$1.22\,pc, derived by maximising the overlap integral of the complementary distributions between the cavities generated by the small and large clouds \citep{2018ApJ...859..166F}. In Fig.~\ref{displace}b, the two components align well after the displacement is applied.

From the spectra presented in Fig.~\ref{exempleline}, we estimate the separation between the LSR velocities of the components to be $\sim$2\,\kms (see Sec.~\ref{morphology}). This separation gives a lower limit to the relative velocity of the clouds. The actual relative velocity might be higher due to projection effects \citep{2015ApJ...807L...4F}. If we apply 45$^{\circ}$ for the viewing angle of the collision to the line of sight (see Sect.~\ref{sec:CCC}), the distance between the two clouds is D\,(pc) = 1.22 /sin(45$^{\circ}$) , and the velocity difference estimated from the observed relative velocity is V\,(\kms) = 2/cos(45$^{\circ}$). The collisional timescale can be estimated to be D/V $\sim$ 0.59\,Myr. If we take into account variable viewing angles of the collision with the line of sight from 30$^{\circ}$ to 60$^{\circ}$, the collision timescale is derived to be 0.35–1.03\,Myr.

\subsection{Alternative kinematic drivers}
\label{drivers} 
We evaluated whether alternative large-scale kinematic drivers could explain the observed velocity structure and found that none provide a consistent explanation comparable to that of CCC. In Section Sect.~\ref{sec:CCC}, we presented observational signatures of a CCC and demonstrated its impact on the large-scale velocity field within the G013.313 region. However, in interpreting observed velocity gradients across clouds and filaments, it is essential to consider alternative mechanisms such as longitudinal collapse, gas inflow, shear motions, rotation, wind-driven acceleration, etc. 

The moment-1 map (Fig.~\ref{G13}c) shows a keyhole-like complementary structure, not the symmetry expected from rotation. A clear rotational axis is also absent, making rotation an unlikely explanation. Although YSOs and clumps are present in G013.313, their impact on the overall velocity gradient is limited. Star-forming activities primarily affect small-scale regions, creating localised turbulence rather than a coherent large-scale gradient. Similarly, the YSO cluster lacks the strong feedback necessary to drive a systematic velocity pattern across the filament. Without a central H{\sc ii} region, expansion cannot account for the observed velocity distribution. Wind-driven acceleration, powered by a nearby stellar cluster, represents a plausible scenario. In Fig.~\ref{G13}a, the 70 $\mu$m emission does not show a prominent compact source. However, it cannot be excluded that some sources are interacting with the surrounding environment, potentially influencing their velocity through expanding winds. However, this interaction is unlikely to significantly affect the overall dynamics of the filament network. \cite{2015MNRAS.447.1059K} showed that shear can produce velocity gradients, but these are usually accompanied by significant deformation and fragmentation of molecular clouds. Therefore, shear-induced velocity gradients are not a plausible primary mechanism.

Although gravitational collapse and longitudinal inflow may explain localised gradients within Hub-1 and Hub-2, they cannot account for the consistent, large-scale velocity field seen across the region. These processes may operate alongside CCC, contributing to material accumulation and collapse in local substructures.

\subsection{HFSs triggered by CCC}
\label{HFS-CCC} 

Building on these observational results, we can infer that the compressive effects driven by CCC may be a key factor in the formation of filamentary structures in this region. The impact of CCC causes local gas to accumulate in the compressed region, forming high-density areas that evolve into filamentary structures \citep{2015MNRAS.453.2471B, 2018PASJ...70S..53I}.

In \cite{2009ApJ...700.1609M}, it is observed that the peak density of a hub is typically much higher than that of its surrounding environment. However, as shown in the upper panels of Fig.~\ref{inflow}, along the filament paths associated with Hub-1 (namely F2–F5 and F4–F3), the column density in the centre of the hub shows little variation, indicating that the density there is similar to its surroundings. This observation implies that Hub-1 may still be in the process of transitioning from a compressed layer to a more concentrated HFS. In contrast, as shown in the bottom panels of Fig.~\ref{inflow}, the intensity distribution near Hub-2, located closer to the G013.313-red component, displays features more consistent with the density enhancements expected in a hub, although further investigation is necessary to confirm this trend. 

In addition, the observed complex inflow motions, combined with the characteristics of CCC and the smooth large-scale distribution of filament velocities and column densities throughout the region, suggest that CCC processes are likely responsible for triggering the formation of the HFS. Furthermore, the presence of local V-shaped velocity structures around intensity peaks, as shown in Fig.~\ref{inflow}, may indicate the existence of local collapse centres, further supporting the ongoing dynamical evolution in this region.

In its current state, G013.313 appears to be relatively young. Its 8\,$\mu$m extinction, combined with the overall gravitationally unbound nature of the region (see Sect.~\ref{SF}), suggests that it is still transitioning from a compressed layer to a HFS. Some theoretical models indicate that the free fall process is highly non-linear \citep{2019MNRAS.490.3061V, 2014ApJ...793...84Z}, progressing slowly in the initial stages and then accelerating dramatically as it approaches completion, at which point hierarchical and chaotic gravitational collapse begins to dominate. This, in turn, implies that the more mature HFSs observed elsewhere may represent the end products of star-forming regions that have evolved to a more advanced stage.

\subsection{Star and cluster formation}
\label{SF} 
Given this connection between CCC and HFS formation, understanding their role in massive star formation becomes crucial. The gravitationally unstable conditions created by CCC not only facilitate hub formation but also enhance accretion flows along filaments, fuelling the growth of dense cores \citep{2018ARA&A..56...41M}. Observations of YSOs and embedded clumps suggest that CCC-induced HFSs may serve as efficient sites for early-stage massive star formation \citep{2019A&A...629A..81T}.

\subsubsection{Star formation potential of clumps}
\label{SF-clump} 
\begin{figure}[h!]
\vspace*{0.2mm}
\centering
\includegraphics[width=0.4\textwidth]{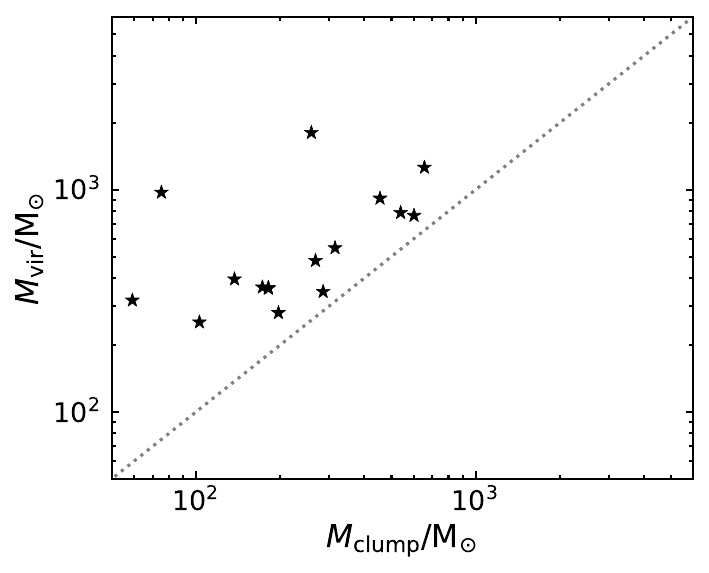}

\includegraphics[width=0.4\textwidth]{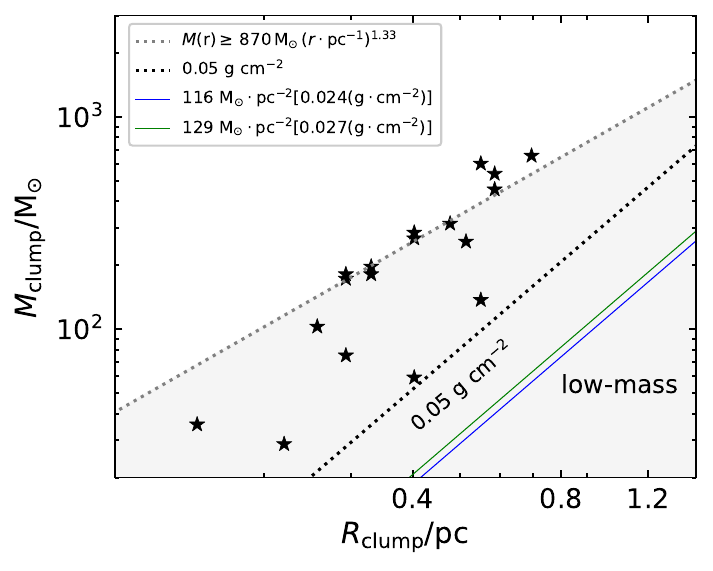}

\includegraphics[width=0.4\textwidth]{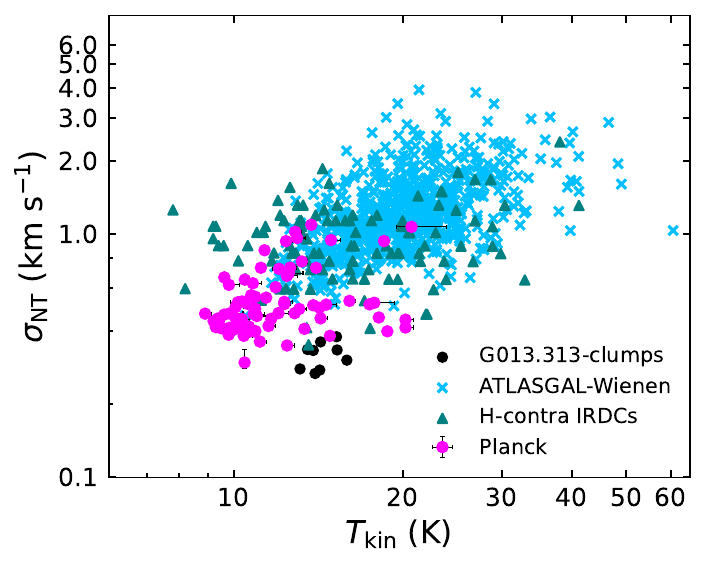}

\caption[]{Top panel: Relation graph between $M_{\rm clump}$ and $M_{\rm vir}$. The dashed black line indicates the $M_{\rm clump}$ = $M_{\rm vir}$ relation. Middle panel: Clump mass as a function of radius for clumps. The grey area indicates the extent of the low-mass star-forming region. The dotted grey line indicates the threshold for the high-mass star formation  $M(\text{r})\geq\,870\,\text{M}_{\odot}\, (r\cdot\text{pc}^{-1})^{1.33}$
determined by \cite{2010ApJ...723L...7K}. The dotted black line indicates the threshold (0.05 g cm$^{-2}$) for the high-mass star formation determined by \cite{2014MNRAS.443.1555U}. The solid blue and green lines at the bottom right represent surface density thresholds for 'efficient' star formation of 116 $\text{M}_{\odot}\cdot \text{pc}^{-2}\left[0.024\left(\text{g}\cdot \text{cm}^{-2}\right)\right]$\citep{2010ApJ...724..687L, 2015A&A...576L...1L} and 129 $\text{M}_{\odot}\cdot \text{pc}^{-2}\left[0.027\left(\text{g}\cdot \text{cm}^{-2}\right)\right]$\citep{2010ApJ...723.1019H}. In the plot, data points for clumps are depicted as black stars. Bottom panel: Non-thermal velocity dispersion $\sigma_{\rm NT}$ vs gas kinetic temperature derived from \NH\,(1,1) and (2,2).}
\label{fig:clump}
\end{figure}

We assess whether the clumps in G013.313 have the physical conditions to form massive stars. In the upper panel of Fig.~\ref{fig:clump}, we present the distribution of $M_{\rm clump}$ versus $M_{\rm vir}$ for the clumps. The dashed line delineates the threshold of gravitational stability. Clumps with ratios above unity are likely unbound, whereas those below that may be prone to instability and collapse. From the graph, it is evident that all identified clumps within the G013.313 region lie above the dashed line. This suggests that these clumps might not be gravitationally bound and implies that their gravitational force is insufficient to counteract turbulence or thermal pressure, potentially leading to expansion or dispersal into the ambient interstellar medium. 

In the middle panel of Fig.~\ref{fig:clump}, we illustrate the relationship between the mass and the effective radius of the clumps, explaining their characteristics. The clumps in G013.313 exhibit mass surface densities significantly exceeding the threshold for active star formation, as proposed by \cite{2010ApJ...724..687L,2015A&A...576L...1L,2010ApJ...723L...7K}. According to the size-constrained mass of the high-mass star formation determined by \cite{2010ApJ...723L...7K}, nine clumps exceed this limit. In contrast to the surface density threshold (0.05\,g \cmm) proposed by \cite{2014MNRAS.443.1555U}, all the clumps in our region exceed this threshold. Hence, the clumps in our region are likely potential sites for high-mass star formation.

The clumps are supported against their gravity by both thermal and non-thermal motions \citep{2003RMxAC..15..293C}. The former is a manifestation of the kinetic temperature within a clump, while the latter originates from star-forming activities such as infall motions and outflows that can broaden the non-thermal velocity dispersion. When turbulent energy is converted into heat, a correlation is expected to exist between kinetic temperature and linewidth \citep{1996A&A...308..573M, 2013A&A...550A.135A, 2016A&A...586A..50G, 2016A&A...595A..94I, 2017A&A...598A..30T, 2018A&A...609A..16T, 2018A&A...611A...6T, 2021A&A...655A..12T}. In this study, we discuss the existence of a correlation between the temperature and turbulence of the clumps in G013.313 traced by dense gas. To achieve this, we utilised the NH$_3$\,(1,1) and (2,2) line ratio derived kinetic temperature and NH$_3$ non-thermal velocity dispersion ($\sigma_{\rm NT}$) as a decent approximation proxy for turbulence (see Sect.~\ref{sec-dispersion} of Appendix.~\ref{NH3}). The non-thermal velocity dispersions versus the kinetic temperature of the sources are shown in the lower panel of Fig.~\ref{fig:clump}. Our results indicate a weak correlation between non-thermal velocity dispersion and kinetic temperature, suggesting that turbulent heating alone may not be the dominant heating mechanism in this region. Given the evidence for CCC and the formation of HFS, shock heating induced by CCC could contribute to the observed gas temperatures. However, more detailed future observations, potentially with higher resolution facilities, could provide deeper insights into the role of CCC-induced shocks and other potential heating sources in this region.

To obtain comprehensive statistics on these clumps, we conducted a comparative analysis of the non-thermal velocity dispersions and kinetic temperatures. Including high-mass star-forming clumps from the ATLASGAL survey \citep{2012A&A...544A.146W}, high contrast infrared dark clouds (IRDCs) \citep{2013A&A...552A..40C}, as well as Planck cold cores \citep{2024A&A...684A.144B}. From the distribution shown in the bottom panel of Fig.~\ref{fig:clump}, we observe that the non-thermal velocity dispersion of our sources is lower than that of the clumps traced by ammonia in ATLASGAL sources and those found in infrared dark clouds. The temperatures of our clumps are generally in agreement with those reported in these studies. In contrast, compared to the Planck cold cores, our clumps appear to be warmer.

In our study of clumps in the G013.313 region, we observed diverse line profile morphologies that likely reflect the combined influence of star formation (e.g., infall and outflow) and larger-scale dynamics induced by CCC and HFS processes. For example, clumps near hub-1 (B2877 and B2881) show blue profiles and B2882 shows a pedestal profile, whereas clumps near hub-2 (B2858 and B2865) also show blue profiles, indicating active inflow, possibly driven by filamentary accretion and gravitational collapse \citep{2015MNRAS.453.2471B}. Although H {\sc ii} regions are not detected in G013.313, the general influence of CCC can induce turbulence and gas dispersal \citep{2013ApJ...774L..31I}, further complicating the line profiles. The ambiguous classification of several clumps underscores the complexity of this region and highlights the need for higher resolution observations to disentangle the overlapping processes.

\subsubsection{Triggered cluster formation by CCC}
\label{SF-yso} 

The spatial distribution of YSOs in the G013.313 region provides insight into the star formation history and the dynamical evolution of the clouds. As described in Sect.~\ref{YSO}, most YSOs are concentrated near the interface between G013.313-blue and G013.313-red. This clustering coincides with the location of the compressed layer traced by BGPS clumps, implying that the CCC between the two clouds has enhanced the formation of dense structures and stellar populations.

The peak hydrogen molecular column densities of G013.313-blue and G013.313-red reach 1.9 and 1.5\,$\times$\,10$^{23}$\,\cmm, respectively. These values exceed the typical threshold for single O-star formation ($\sim$10$^{22}$\,\cmm) and approach or surpass the threshold for forming multiple O-stars ($\sim$10$^{23}$\,\cmm) as outlined in \cite{2021PASJ...73S...1F}. The column density in the compressed region, reaching approximately 2\,$\times$\,10$^{23}$\,\cmm, lies between the values observed in well-known massive star-forming environments such as RCW38 \citep{2016ApJ...820...26F} and M20 \citep{2011ApJ...738...46T,2017ApJ...835..142T}. This supports the hypothesis that the region hosts conditions favourable for forming at least one O-type star along with a surrounding population of lower-mass stars. Importantly, no nearby H{\sc ii}  regions or SNRs are detected in this area that could otherwise explain the local compression. Thus, CCC emerges as the most plausible external driver of the high column density and triggered star formation.

Further support for the collision-triggered scenario comes from the stellar age estimates. The typical ages of Class I and II YSOs (0.4--2 Myr; \citealt{2009ApJS..181..321E, 2015ApJS..220...11D, 2020MNRAS.499.3620I}) align well with the estimated collision timescale of 0.35--1.03 Myr. This temporal consistency reinforces the interpretation that the observed YSO population and clump formation were likely initiated by the recent collision event. The higher YSO concentration in Hub-1 compared to Hub-2 may reflect differences in the local efficiency of star formation or varying stages of evolution within the collision interface.

\section{Summary}
\label{sect:summary}
We made observations of \co, \xco, \yco, and \hco~\jxs with the PMODLH 14 m telescope, complemented by data from the Nanshan 26 m radio telescope for the inversion transitions \NH\,(1,1) and (2,2) in the G013.313+0.199 region. The key findings of this study are as follows.

\begin{enumerate}
\item
We present observational data and analyses of molecular gas dynamics using CO line observations, highlighting several key signatures of CCC, such as complementary distributions, U- and V-shaped features, and bridge structures. These features suggest dynamic interactions between molecular clouds, supporting the hypothesis that CCCs play a significant role in shaping the physical and kinematic properties of the region.

\item We determine that the masses of the molecular clouds G013.313-blue and G013.313-red were approximately 0.86 and 1.10\,$\times$\,10$^{5}$\,\msun, with peak molecular hydrogen column densities of 1.9 and 1.5\,$\times$\,10$^{23}$. Our analysis of the ammonia spectral lines reveals that the blue-shifted cloud has a higher ammonia column density, suggesting higher mass concentrations in dense regions. Both clouds exhibit high Mach numbers, indicating that non-thermal processes significantly influence their dynamics.

\item
We identify 21 BGPS clumps within G013.313, which serve as potential sites for high-mass star formation. Although these clumps are not gravitationally bound, their virial mass and gas mass distributions approach the critical line where the two masses equalise. Furthermore, the observed correlation between clump size and mass, along with temperature and non-thermal velocity dispersions, suggests an active star-formation potential that may influence the future evolution of these clumps.

\item
Additionally, we identify 94 YSOs within G013.313, with a notable concentration of Class I and II YSOs indicative of early stellar evolution stages. Their spatial distribution aligns with the structures formed by cloud interactions, further supporting the hypothesis of collision-induced star formation.
\item
We estimate a collision timescale of 0.35 to 1.03\,Myr, which aligns with the ages of the identified YSOs, providing a temporal context for the ongoing star formation processes.  This temporal alignment, along with the observed physical conditions, further strengthens the hypothesis that CCCs are responsible for the formation of dense clumps and YSO clusters in G013.313.

\item 
We also considered alternative kinematic drivers, such as rotation, wind-driven acceleration, and gas inflow, ruling them out as primary mechanisms. The observed velocity gradients and filamentary structures in the region align with the effects of CCC, which may induce gas compression and gravitational collapse, leading to the formation of stars. This study highlights the importance of CCC in the evolution of molecular clouds, the formation of HFSs, and subsequent star formation processes in the G013.313 region.

\item 
Furthermore, we observe that CCCs are probably the driving force behind the formation of HFSs in the region. These structures, which are crucial for the formation of massive stars, are seen to align with regions of intense compression resulting from the cloud–cloud interactions. This supports the notion that CCCs influence not only star formation but also directly contribute to the formation of the dense cores required for massive star formation.

\item
This study demonstrates that the G013.313 region is a compelling example of how CCCs can facilitate star formation, emphasising the need for further observations to understand the implications of these interactions on the formation of massive stars and star clusters in similar environments.
\end{enumerate}

\begin{acknowledgements}

This work was mainly funded by the National Key R\&D Program of China under the grant (Nos.\,2022YFA1603103 and 2023YFA1608002). It was also partially funded by the Regional Collaborative Innovation Project of Xinjiang Uyghur Autonomous Region grant No.\,2022E01050, the National Natural Science Foundation of China under grant Nos.\,12173075, 11973076, 12103082, and 12403033, the Natural Science Foundation of Xinjiang Uygur Autonomous Region under grant No.\,2022D01E06 and 2023D01A11, the Chinese Academy of Sciences (CAS) “Light of West China” Program under grant Nos.\,2021-XBQNXZ-028 and XBZG-ZDSYS-202212, the Xinjiang Key Laboratory of Radio Astrophysics under grant No.\,2022D04033. The Tianshan Talent Training Program of the Xinjiang Uygur Autonomous Region under grant No.\,2022TSYCLJ0005, the Youth Innovation Promotion Association CAS, the Tianshan Talent Training Program and Tianchi Talent Project of Xinjiang Uyghur Autonomous Region. C.\,Henkel has been funded by the Chinese Academy of Sciences President's International Fellowship Initiative grant No.\,2025PVA0048. In addition, this research has been founded by the Science Committee of the Ministry of Science and Higher Education of the Republic of Kazakhstan grant No.\,AP26100637.  This work is based on observations made with the Nanshan 26-m Radio Telescope and is partly supported by the Operation, Maintenance, and Upgrading Fund for Astronomical Telescopes and Facility Instruments, funded by the Ministry of Finance of China (MOF) and administered by the Chinese Academy of Sciences (CAS), as well as the Urumqi Nanshan Astronomy and Deep Space Exploration Observation and Research Station in Xinjiang. We extend our gratitude to the staff of the Purple Mountain Observatory in Delingha, particularly our observational colleagues, for their contributions in acquiring the high-quality observations. We thank the anonymous referee for the constructive comments that helped improve the quality of this manuscript.
\end{acknowledgements}
\bibliography{sample63}
\bibliographystyle{aa}

\begin{appendix}

\section{Derived CO parameters}
\label{appendA}

Our calculations assume that the region in a state of local thermodynamical equilibrium (LTE) and the $^{13}$CO emission is optically thin. Adopting a beam-filling factor of unity, the excitation temperature for CO is derived from the equation \citep{2010ApJ...721..686P,2015ApJ...805...58K}
\begin{equation}\label{}
\begin{split}
T_{\rm ex (CO)} =5.33 \left\{ \rm ln \left[1+ \frac{5.33}{\left(T_{\rm mb,^{12}\rm CO}+0.818\right)}\right] \right\}^{-1}\, \rm K,
\end{split}
\end{equation}  
where $T_{\rm mb,^{12}\rm CO}$ is the peak intensity of $^{12}$CO \jxs in units of K. We then obtain the optical depths, $\tau$, and the column densities, $N_{\rm ^{13}CO}$, $N_{\rm C^{18}O}$ for the $^{13}$CO and C$^{18}$O molecules using the following equations \citep{1994ApJ...429..694L,1998ApJS..117..387K}.
\begin{equation}\label{}
	\begin{aligned}
		\tau_{\rm ^{13}CO} & =-\ln \left\{1 - \frac{T_{\rm mb,^{13}CO}}{5.29([\exp(5.29 / T_{\rm ex (CO)}-1]^{-1}-0.164)}\right\}\,,
	\end{aligned}
\end{equation}

\begin{equation}\label{}
	\begin{aligned}
		\tau_{\rm C^{18}O} & =-\ln \left\{1 - \frac{T_{\rm mb,C^{18}O}}{5.27([\exp(5.27 / T_{\rm ex (CO)}-1]^{-1}-0.166)}\right\}\,,
	\end{aligned}
\end{equation}

\begin{equation}\label{}
	\begin{aligned}
		N_{\rm ^{13}CO} & =2.42\times 10^{14} \frac{\tau_{\rm ^{13}CO} \Delta v_{\rm ^{13}CO} T_{\rm ex (CO)}}{1-\exp(-5.29/T_{\rm ex (CO)})}\, \rm cm^{-2},
	\end{aligned}
\end{equation}
and 
\begin{equation}\label{}
	\begin{aligned}
		N_{\rm C ^{18}O} & =2.42\times 10^{14} \frac{\tau_{\rm C^{18}O} \Delta v_{\rm C^{18}O} T_{\rm ex (CO)}}{1-\exp(-5.27/T_{\rm ex (CO)})}\, \rm cm^{-2},
	\end{aligned}
\end{equation}

where $T_{\rm mb,^{13}\rm CO}$ and $T_{\rm mb,\rm C^{18}O}$ are peak intensities in K, while $\Delta v_{\rm ^{13}CO}$ and $\Delta v_{\rm C^{18}O}$ are the line widths in\,km\,s$^{-1}$. In the calculation, the excitation temperatures of \xco~ have been assigned equal to those of $^{12}$CO. Molecular hydrogen column densities were calculated by the fractional abundances of [H$_{2}$]/[$^{13}$CO] = 89\,$\times$\,10$^{4}$ \citep{1980ApJ...237....9M} and [H$_{2}$]/[C$^{18}$O] = 7\,$\times$\,10$^{6}$ \citep{1982ApJ...262..590F} in the solar neighbourhood. 

The hydrogen column density ($N\rm _{H_{2}}$) values derived from $^{13}$CO and C$^{18}$O are quite close to each other, suggesting that both $^{13}$CO and C$^{18}$O are optically thin and the optical depth effect can be ignored \citep{2013ApJS..209...37M}. Therefore, we used $^{13}$CO to calculate $N\rm _{H_{2}}$, as described in Sect.~\ref{parameters} and shown in Fig.~\ref{calculate} of Appendix.~\ref{appendA}. 

The mass can be calculated by aggregating the masses of all the pixels within each cloud and clump. The mass for each pixel is determined by
\begin{equation}\label{Massgas}
\begin{split}
M = 2\,N\rm _{H_{2}}(^{13}CO)\,\mu_{\rm H_{2}}\,m_{\rm H}\,A_{\rm pixel},
\end{split}
\end{equation}  
where $\mu_{\rm H_{2}}$ = 2.72 is the mean molecular weight \citep{2010A&A...513A..67B}, $m_{\rm H}$ = 1.67\,$\times$\,10$^{-27}$ kg is the mass of a hydrogen atom, and $A_{\rm pixel}$ is the area of each pixel within the cloud and clump.

The virial mass can be calculated from the average velocity dispersion \citep{2021A&A...650A.164M}:
\begin{equation}\label{virial}
\begin{split}
M_{\rm vir} = \frac{5\sigma_{\rm v}^{2}R_{\rm eff}}{G},
\end{split}
\end{equation}  
where $\sigma_{\rm v}$ is the velocity dispersion of the clump and $G$ is the gravitational constant \citep{1988ApJ...333..821M}. 

We also estimate the virial stability of these clumps by calculating the virial parameter, $\alpha$ =  ${M_{\rm vir}}/{M_{\rm clump}}$. We assume that these clumps follow a spherical density distribution. In the absence of pressure supporting the clump, $\alpha$ < 1 means that the clump is gravitationally unstable and collapsing, while $\alpha$ > 2 suggests that the kinetic energy is higher than the gravitational binding energy and that the clump is dissipating. A value of $\alpha$ between 1 and 2 is interpreted as an approximate equilibrium between gravitational binding and kinetic energies.

\section{PV diagram and channel Maps}
\label{appendB}

\begin{figure}[h]
\vspace*{0.2mm}
\centering
\includegraphics[width=0.45\textwidth]
{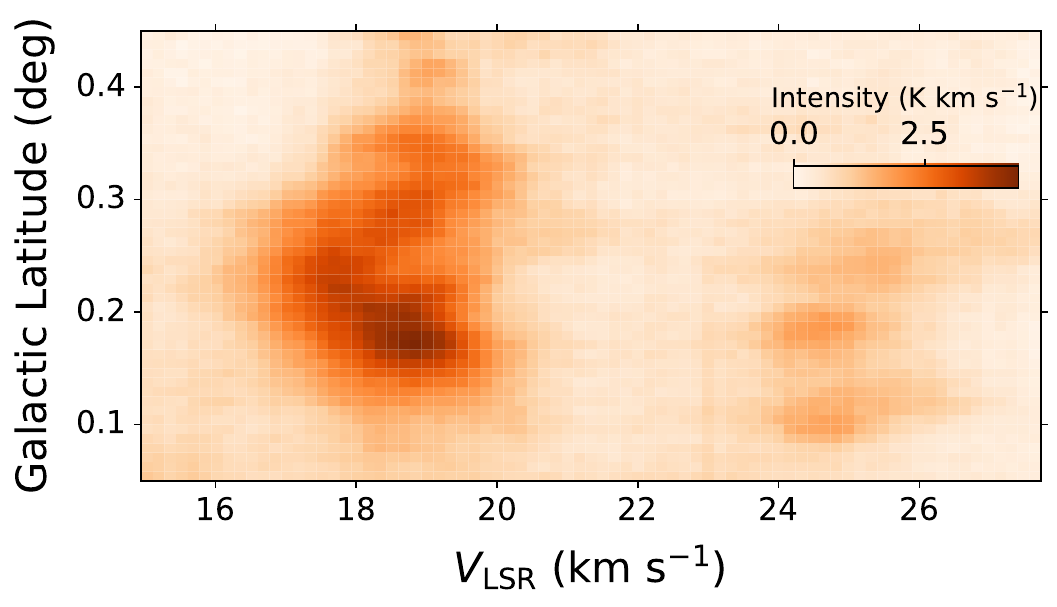}
\caption[]{\xco~position-velocity diagram for the velocity range of 15-27~\kms, with latitude-velocity data integrated across a longitude range of \(l=13.06^\circ \) to \(l=13.54^\circ \). Integration ranges are indicated in the upper right corner.}
\label{pv}
\end{figure}

\begin{figure}[h]
\vspace*{0.2mm}
\centering
\includegraphics[width=0.49\textwidth]
{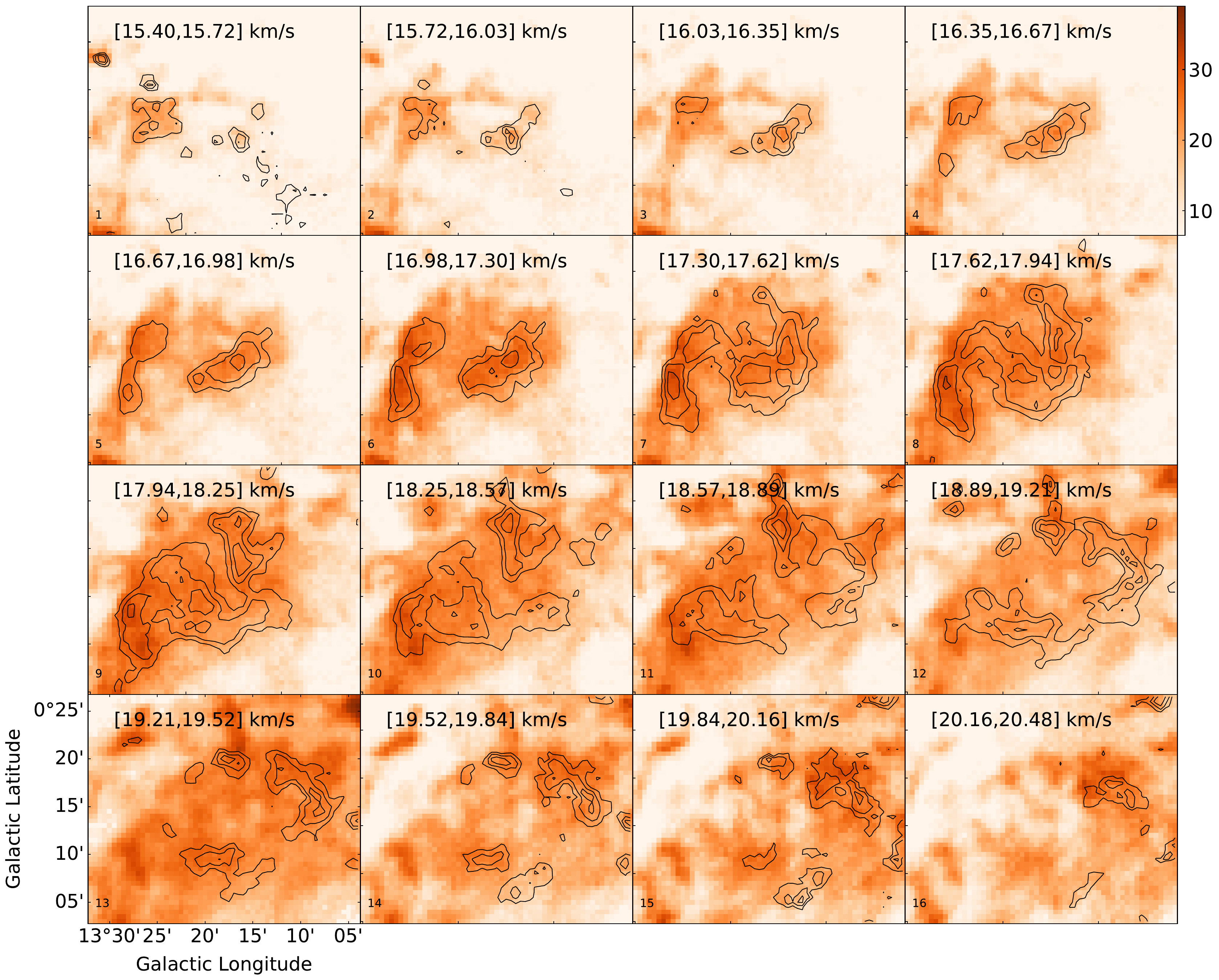}
\caption[]{\co~channel maps (background) with overlaid \xco~contours at the levels of 45\%–85\% of the peak intensity (with steps of 20\%). The velocity range is displayed in the upper left corner of each panel.}
\label{channel-co}
\end{figure}

\begin{figure}[h]
\vspace*{0.2mm}
\centering
\includegraphics[width=0.49\textwidth]
{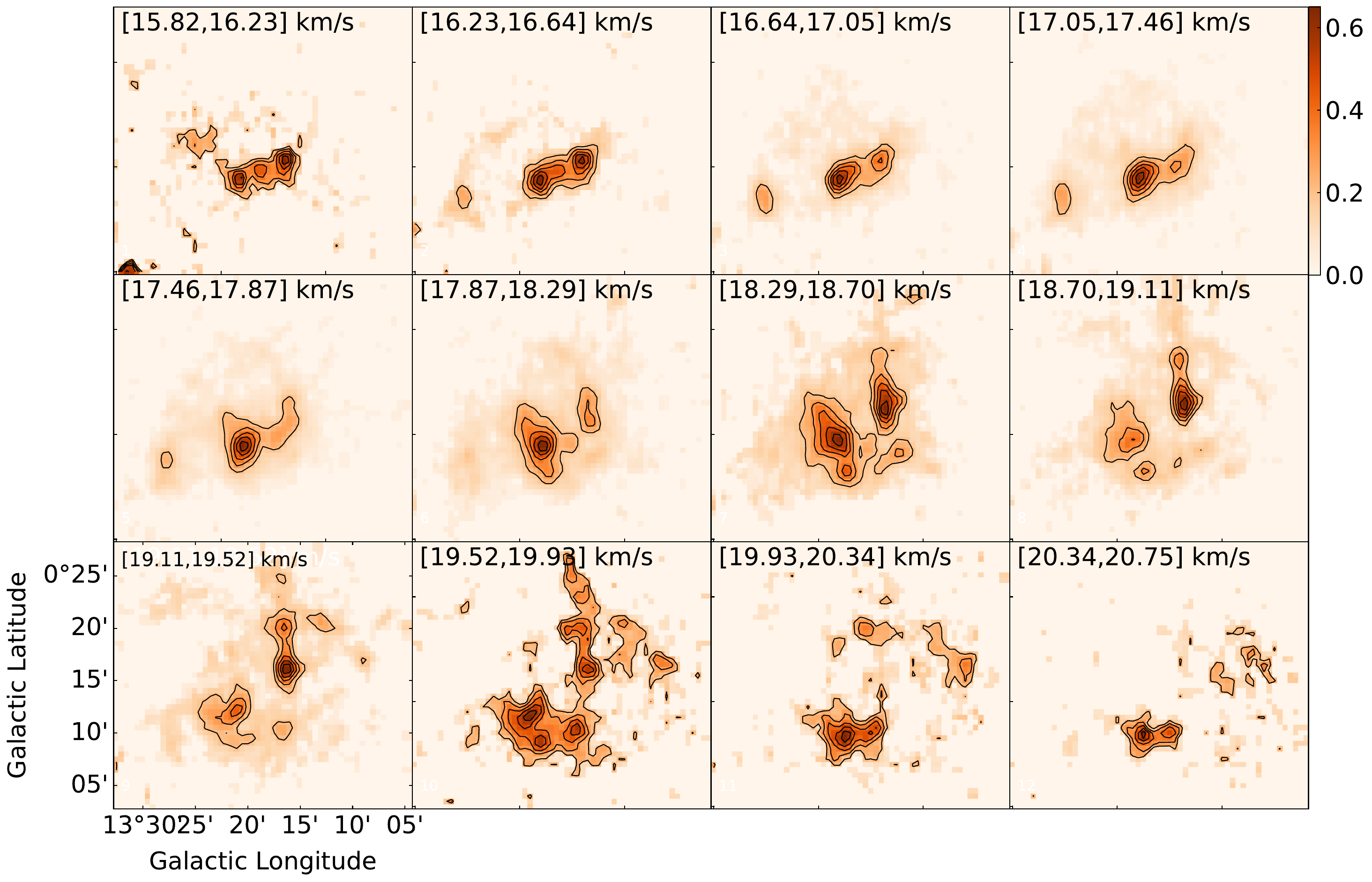}
\caption[]{\hco~channel maps with the steps of 0.31\,\kms and overlaid \xco~contours at levels of 30\%–90\% of the peak intensity (with steps of 15\%). The velocity range is displayed in the upper left corner of each panel.}
\label{channel-hco}
\end{figure}

\section{Filfinder algorithm}\label{finfinder}

We identified filamentary structures in the molecular cloud using Spitzer IRAC 8\,$\mu$m emission and \xco~(1–0) data within the velocity range of 17.93–18.76\,\kms. This velocity range lies between the two main velocity components (centred at $\sim$17.6 and $\sim$19.5\,\kms)(see Sect.~\ref{kinematic}). In the CCC scenario, this intermediate velocity range is expected to trace the shock-compressed interface layer formed by the collision, where filamentary structures and dense cores are often observed to develop \citep[e.g.][]{2018PASJ...70S..53I, 2017ApJ...835..142T}. Therefore, we selected this velocity interval to isolate potential CCC-induced structures while minimising confusion from unrelated foreground and background emissions. Similar approaches using intermediate-velocity filament identification based on molecular line emission and kinematic continuity have also been adopted in other candidates for CCC \citep[e.g.][]{2024MNRAS.528.2199R}, further supporting the relevance of this method.

\begin{figure}[h!]
\vspace*{0.2mm}
\centering
\includegraphics[width=0.4\textwidth]{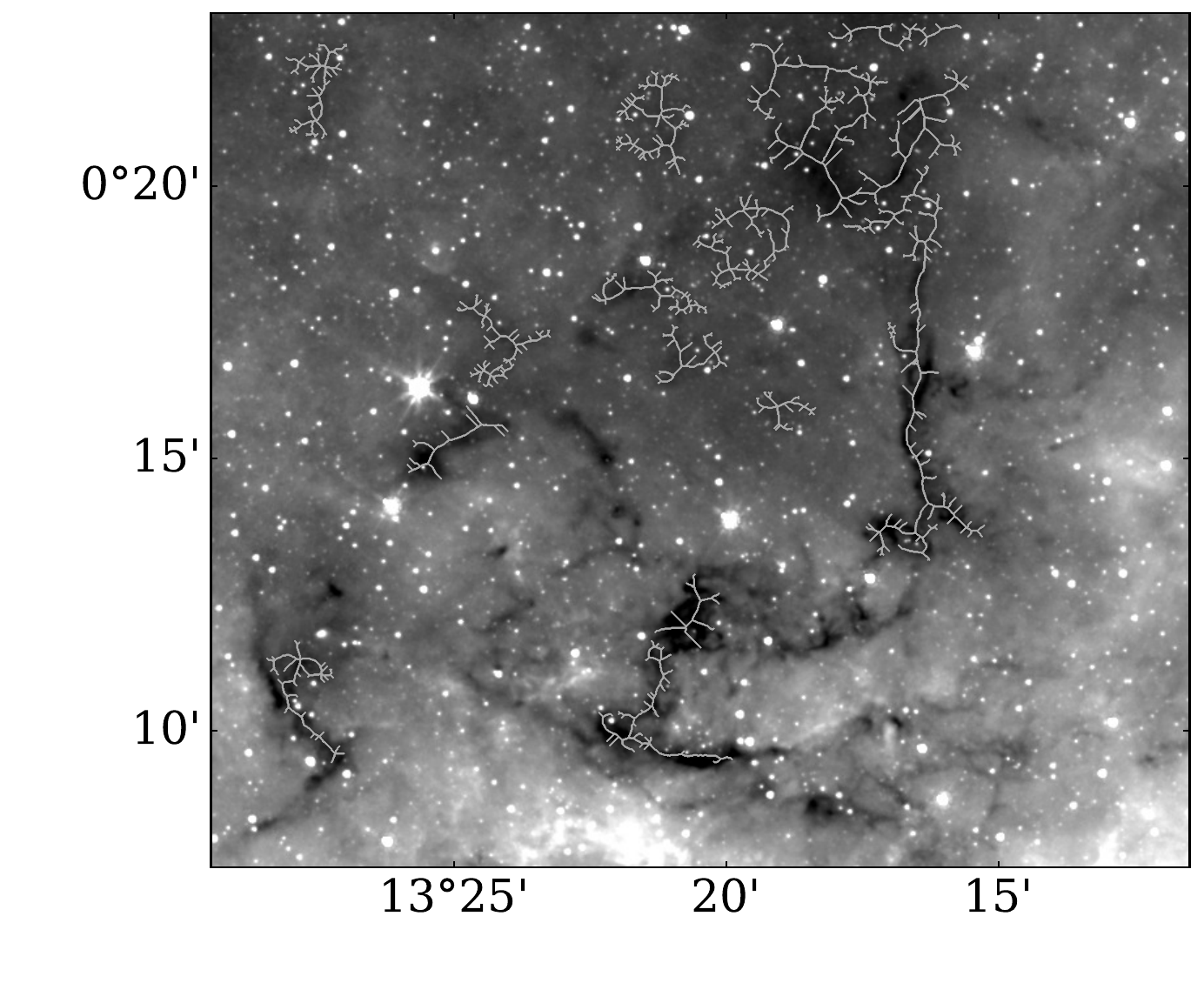}

\includegraphics[width=0.4\textwidth]{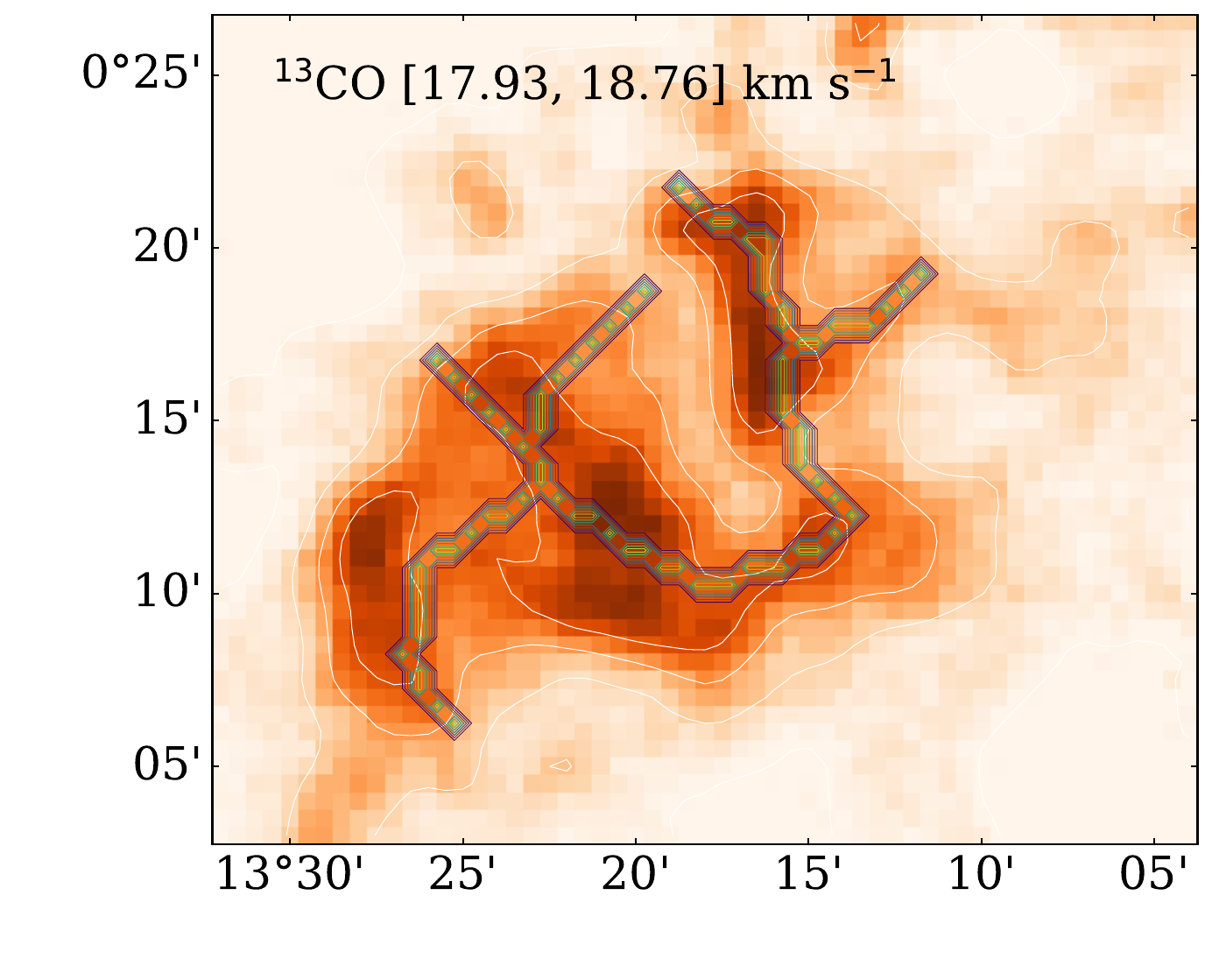}

\includegraphics[width=0.4\textwidth]{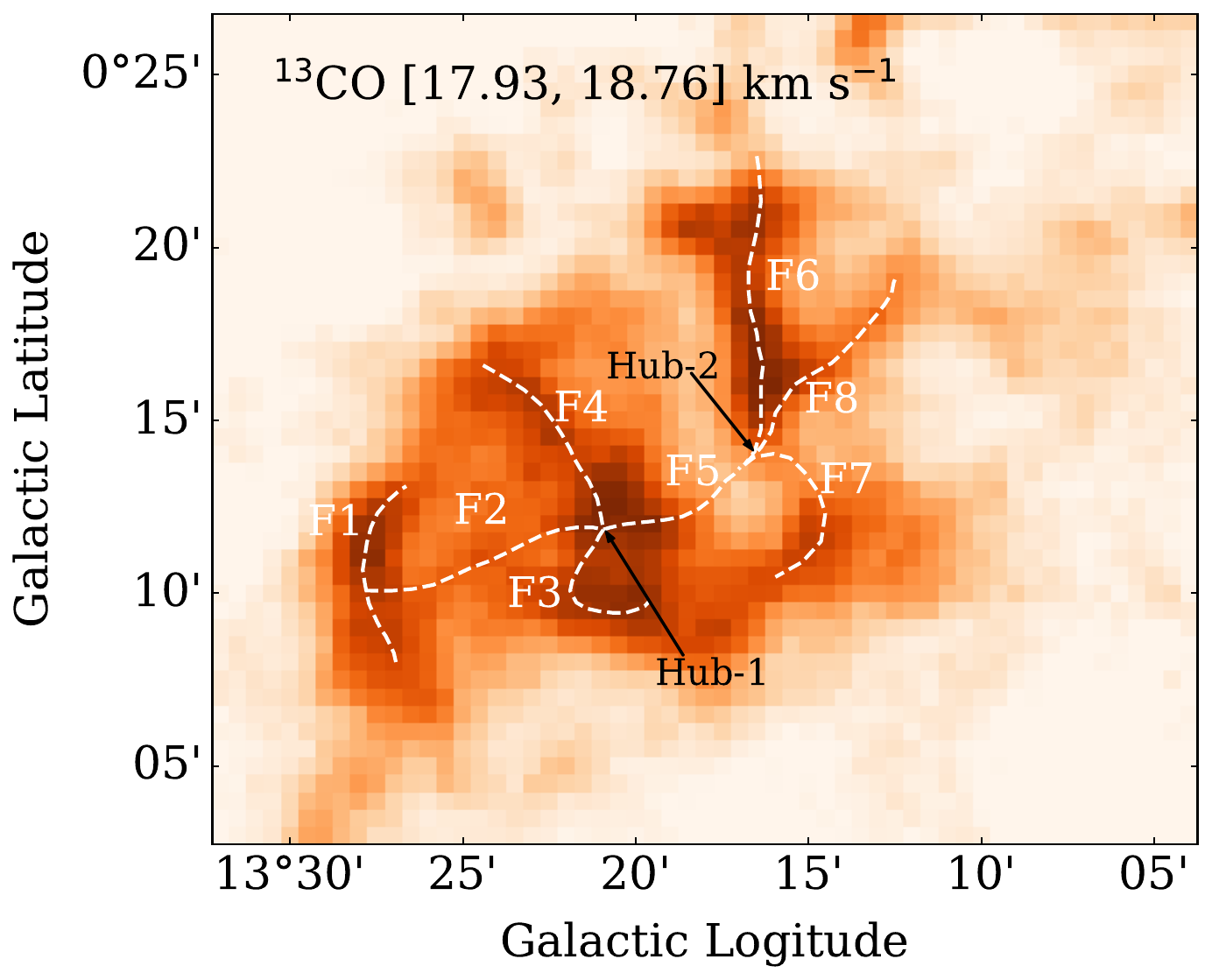}
\caption[]{Filamentary structures identified from Spitzer 8\,$\mu$m and \xco(1-0) data. Top Panel: Filamentary structures identified using Spitzer IRAC 8\,$\mu$m emission with the FilFinder algorithm. Middle Panel: Filamentary structures identified using \xco~(1-0) data in the velocity range of 17.93–18.76\,\kms. Bottom Panel: Manually refined filament paths overlaid on the \xco(1-0) integrated intensity map (17.93–18.76\,\kms), with HFS-1 and HFS-2 marked.}
\label{fig:filfinder}
\end{figure}

Filamentary structures in this region were identified by applying the Python FilFinder package \citep{2015MNRAS.452.3435K} separately to the infrared absorption features and the molecular gas distribution. The results of the Spitzer IRAC 8\,$\mu$m image and the \xco~(1–0) integrated intensity map are shown in the top and middle panels of Fig.~\ref{fig:filfinder}. As demonstrated in Fig. 2 of \citet{2022MNRAS.514.6038Z}, extinction features in Spitzer 8\,$\mu$m emission can effectively trace HFSs, supporting our approach.

Since Spitzer 8\,$\mu$m images lack velocity information, and \xco-based filament identification may be affected by line broadening and projection effects within even a narrow velocity range, we further validated our results by examining \xco~ channel maps (see Fig.~\ref{channel-co} in Appendix.~\ref{appendB}). The filamentary structures identified in the intermediate velocity range exhibit spatial continuity across multiple channels and show good morphological agreement with the infrared extinction features, strengthening their reliability. Based on this multi-tracer, kinematically informed approach, we manually determined the final filament paths for further analysis (see the bottom panel of Fig.~\ref{fig:filfinder}).

\section{\NH\,results}\label{NH3}
\subsection{Kinetic temperature}
With the measured data, the rotational temperature ($T_{\rm rot}$), kinetic temperature ($T_{\rm kin}$), NH$_3$ column density ($N_{\rm NH_{3}}$), thermal velocity dispersion $\sigma_{\rm Therm}$, non-thermal velocity dispersion $\sigma_{\rm NT}$, thermal-to-non-thermal pressure ratio $R_{\rm p}$, thermal sound speed $c_{\rm s}$ and Mach number ($\mathcal{M}$) can be calculated.

\label{sect-kinetic_temperature}
Once the optical depth is determined by the 'NH$_3$\,(1,1)' fitting as described in Sect.~\ref{nh3-obser}, we can calculate the excitation temperature of the NH$_3$\,(1,1) inversion transition through the relation \citep{1983ARA&A..21..239H}.
\begin{eqnarray}
\label{Eq1_1}
T_{\mbox{\tiny ex}}=\frac{T_{\rm MB (1,1)} }{\left( 1- \rm exp(-\tau )\right)}\ \ +2.7 \,{\rm K,}
\end{eqnarray}
where $T_{\rm MB}$ and $\tau$ represent the temperature and the optical depth of the (1,1) line derived using the GILDAS built-in 'GAUSS' and 'NH$_3$\,(1,1)' fitting methods.
 
Since the relative populations of the $K$ = 1 and 2 ladders of NH$_3$ are not directly connected radiatively, they are highly sensitive to collisional processes.
This allows us to use them as a thermometer of the gas kinetic temperature. The method described in \cite{1983ARA&A..21..239H}, has been used to obtain the rotation temperature. 

The rotation temperature is given by the expression

\begin{eqnarray}
\label{Eq1}
T_{\mbox{\tiny rot}}=\frac{-41.5}{\ln\left( \frac{-0.282}{\tau}\ln
\left(1-\frac{T_{\rm MB}(2,2)}{T_{\rm MB}(1,1)}\left( 1- \rm exp(-\tau )\right)
 \right) \right)}\ \  {\rm K,}
\end{eqnarray}
where $T_{\rm MB}$ (2,2) is the main beam brightness temperatures of the (2,2) line derived
using the GILDAS built-in 'GAUSS' fitting method.  
 
We estimated the kinetic temperature $T_{\rm kin}$ using the approximation of \citet{2004A&A...416..191T} :
\begin{eqnarray}
\label{Eq2}
T_{\mbox{\tiny kin}}=\frac{T_{\mbox{\tiny rot}}(1,2)}{1-\frac{T_{\mbox{\tiny rot}}(1,2)}
{41.5}\ln\left( 1+1.1 \, {\rm exp}\left(\frac{-16}{T_{\mbox{\tiny rot}}(1,2)}\right)\right)}\ \ {\rm K,}
\end{eqnarray}
where the energy gap between the (1,1) and (2,2) states is $\Delta E_{\rm 12}$= 41.5\,K. This approximation has been derived with Monte Carlo models and should provide an accuracy of 5\% in the range between 5 and 20\,K. Most of our sources can be found in this interval.

\subsection{NH$_3$ column density}
\label{NNH3-cal}
Computing the ammonia column density requires the optical depth and intrinsic line width of the (1, 1) inversion transition along with the rotational temperature, which is obtained from Eqs.\,(\ref{Eq1}). As the optical depth and the rotational temperature depend only on the intensity ratio of the (1,1) and (2,2) lines, the resulting column density is a source-averaged quantity.

Realistically assuming that for our cold sources, the bulk of the ammonia populations resides in the metastable ($J$ = $K$), ($J$,$K$) = (0,0) to (3,3) levels, the total NH$_3$ column densities can be calculated from NH$_3$\,(1,1) following \cite{2004tra..book.....R},
\begin{eqnarray}
\label{Eq3}
N_{\rm {NH_{3}}} \approx N(1,1) \Bigg( \frac{1}{3} \, {\rm exp} \left(\frac{23.1}
{T_{\mbox{\tiny rot}}(1,2)}\right)+1+ \frac{5}{3}\,{\rm exp} \left(-\frac{41.2}{T_{\mbox{\tiny rot}}(1,2)}\right)\\  \nonumber
+\frac{14}{3} \, {\rm exp}\left( -\frac{99.4}{T_{\mbox{\tiny rot}}(1,2)}\right)\Bigg)\ \ {\rm cm^{-2}}
\end{eqnarray}
 and 
\begin{eqnarray}
\label{Eq4}
N(1,1)=6.60\times10^{4}\,\Delta v\,\tau\,\frac{ \,T_{\rm rot}}{\,\nu}\,\ \ {\rm cm^{-2},}
\end{eqnarray}
where $N$(1,1) is the column density of the NH$_3$\,(1,1) transition, the FWHM line width $\Delta v$ is in \,km\,s$^{-1}$, the line frequency $\nu$ is in GHz, and the rotational temperature $T_{\rm rot}$ is in Kelvin. 

\subsection{Velocity dispersions, sound speed, and gas pressure ratio of \NH}
\label{sec-dispersion}
The observed line widths provide a measure of the internal motions within each clump. Here, we computed non-thermal velocity dispersion
($\sigma_{\rm NT}$), thermal velocity dispersion ($\sigma_{\rm TH}$), and sound speed ($c_{\rm s}$) following \cite{2014A&A...567A..78L}, \cite{2017A&A...598A..30T, 2018A&A...609A..16T, 2018A&A...611A...6T, 2021A&A...655A..12T}. 
\begin{eqnarray}
\sigma_{\rm NT} = \sqrt{\sigma_{\rm v}^{2}-\sigma_{\rm TH}^{2}}\ \ \,\rm km\,s^{-1},
\end{eqnarray}
where $\sigma_{\rm v}$ = $\Delta v(1,1)$/8ln(2), and $\sigma_{\rm TH}$ = ($k_{\rm B}$$T_{\rm kin}$/17$m_{\rm H}$)$^{1/2}$. $m_{\rm H}$ is the mass of a single hydrogen atom and $k_{\rm B}$ is the Boltzmann constant. 

The thermal sound speed can be calculated with 
\begin{eqnarray}\label{soundspeed}
c_{\rm s} = \sqrt{\frac{k_{\rm B}T_{\rm kin}}{\mu m_{\rm H}}} \ \ \,\rm km\,s^{-1},
\end{eqnarray}
where $\mu$ = 2.37 is the mean molecular weight for molecular clouds \citep{2016ApJ...819...66D}.

We also calculated the thermal-to-non-thermal pressure ratio
($R_{\rm P}$=$\sigma_{\rm TH}^{2}$/$\sigma_{\rm NT}^{2}$; \citealt{2003ApJ...586..286L})
and Mach number (given as $\mathcal{M}$=$\sigma_{\rm NT}$/$c_{\rm s}$). The mean values for the clouds are summarised in Table~\ref{tab:2parameter}.
The derived values of $T_{\rm kin}$, $N_{\rm NH_{3}}$, $\sigma_{\rm NT}$ and $\mathcal{M}$ for individual clumps are listed in Table~\ref{clumps} of Appendix.~\ref{append-clump}. 

\begin{figure}[h!]
\vspace*{0.2mm}
\centering
\includegraphics[width=0.49\textwidth]
{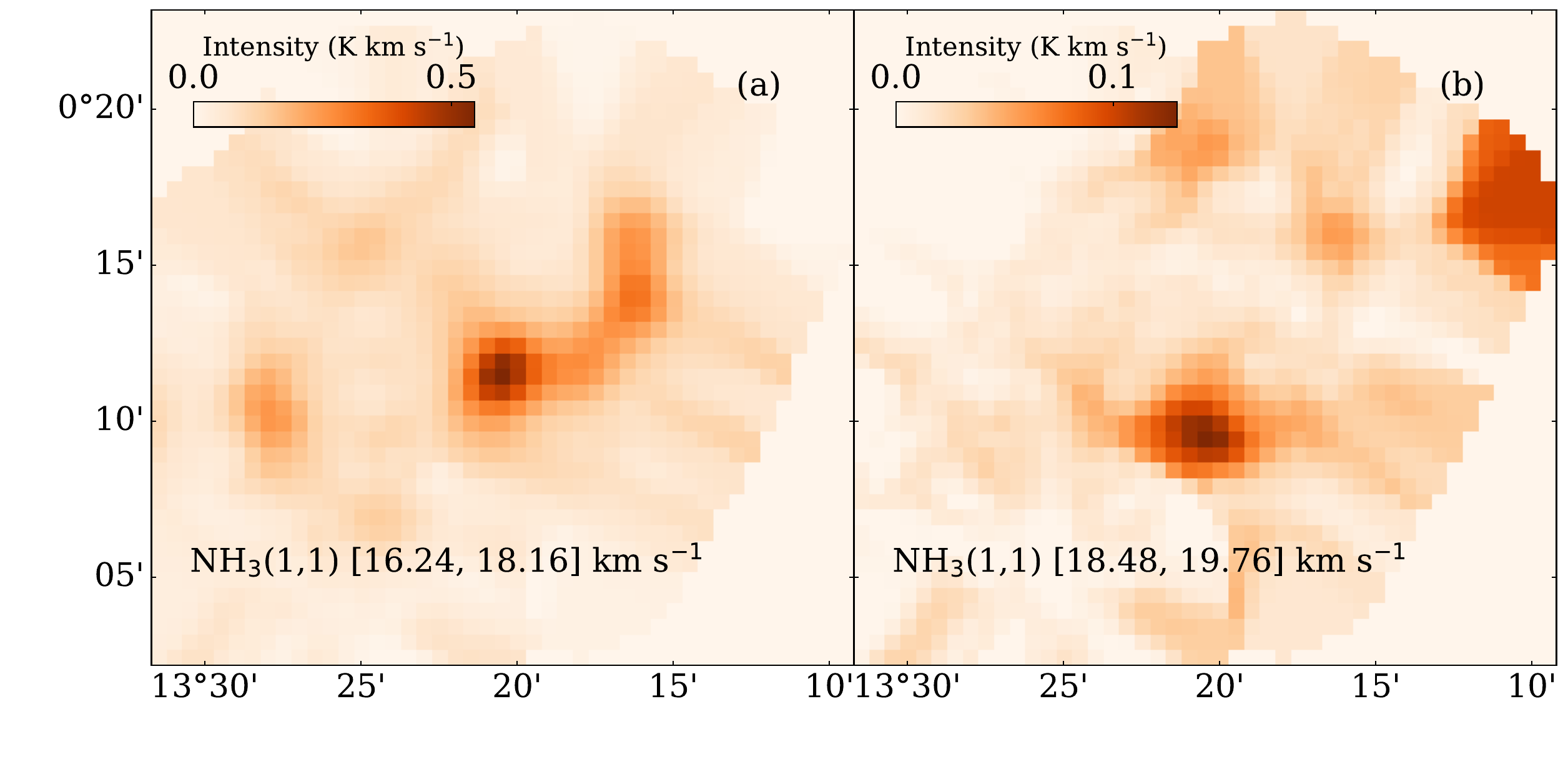}

\includegraphics[width=0.49\textwidth]
{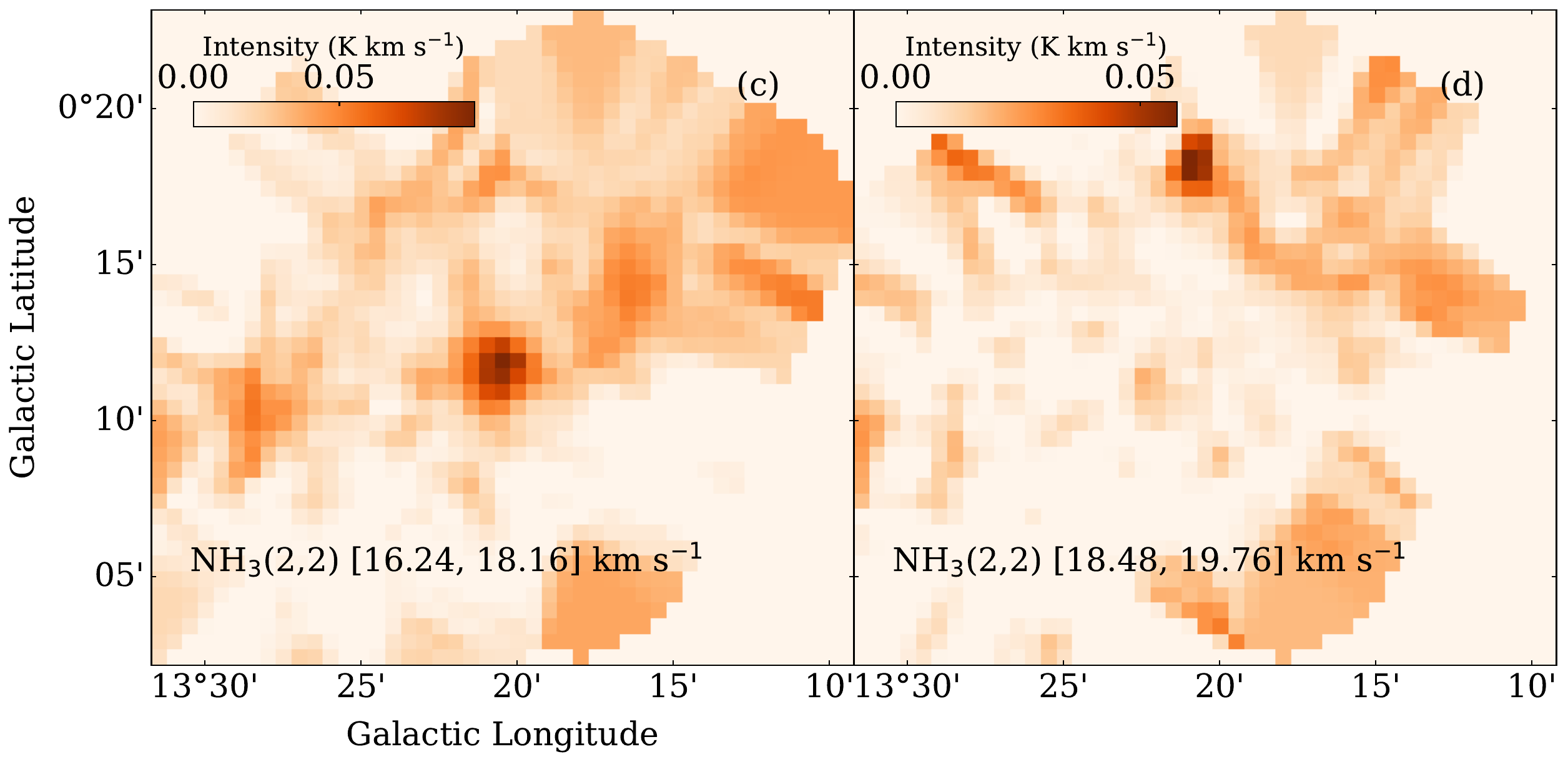}
\caption[]{Integrated intensity maps of the blue and red clouds are displayed in panels (a, b, c, and d). The upper two panels (a–b) illustrate the \NH\,(1,1) emission line, while the bottom two panels (c–d) depict the \NH\,(2,2) emission line.}
\label{nh3-moment0}
\end{figure}

\twocolumn
\section{Clumps parameters}
\label{append-clump}

\label{appendixF}

\begin{figure}[h]
\vspace*{0.2mm}
\centering
\includegraphics[width=0.4\textwidth]
{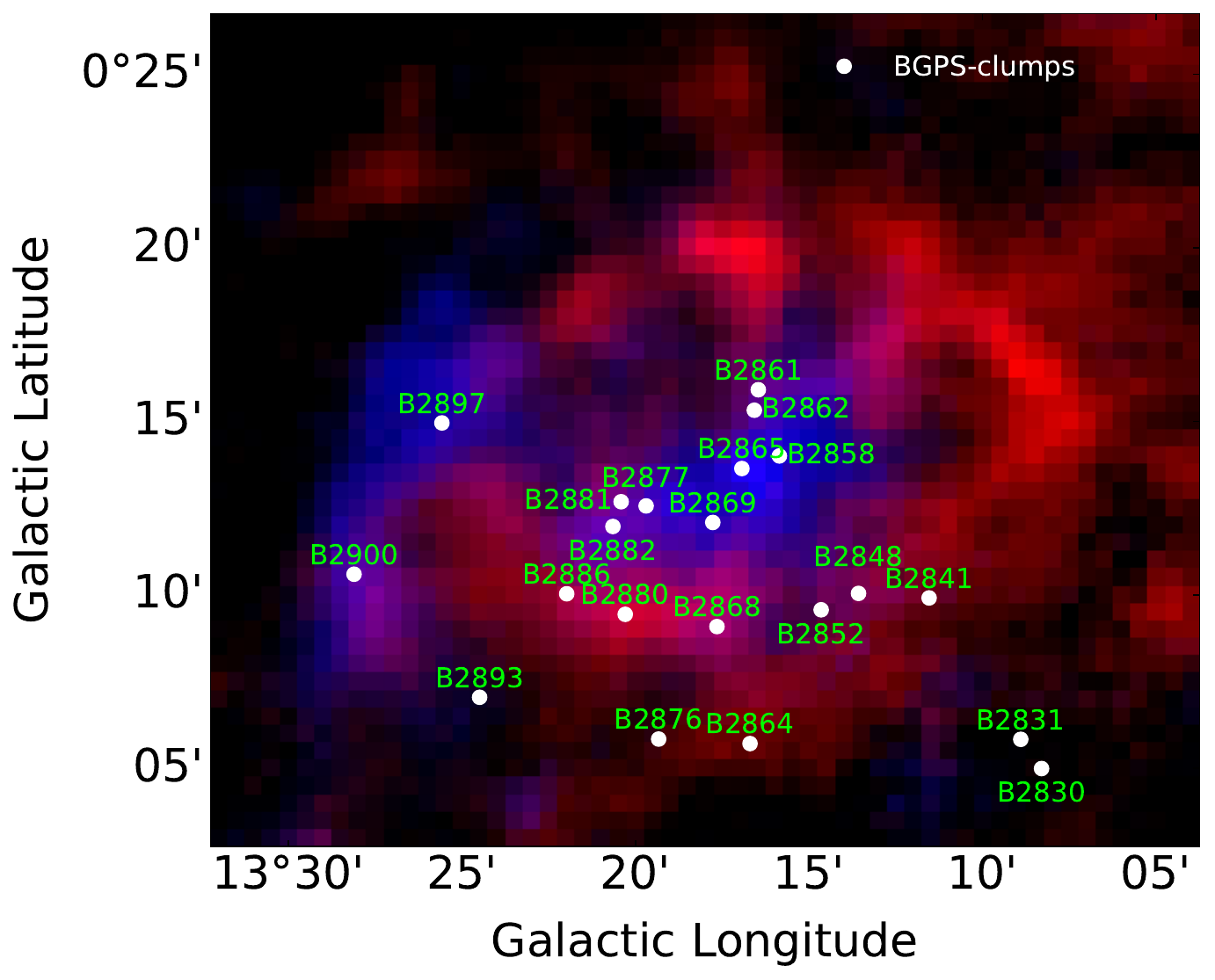}

\includegraphics[width=0.49\textwidth]
{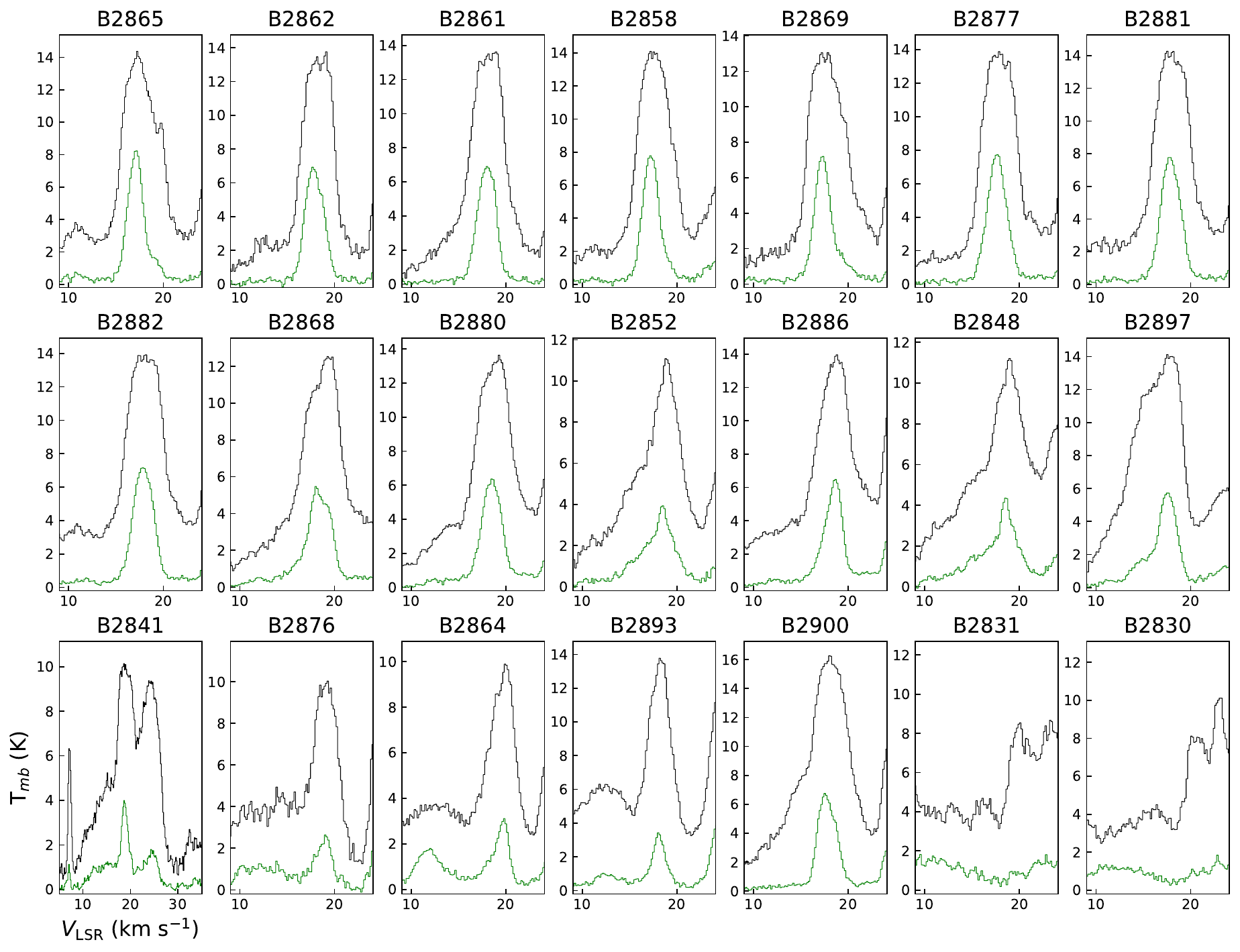}
\caption[]{Top panel: Two-colour composite image of G013.313 employs \xco~data, with blue indicating the blue-shifted cloud and red representing the red-shifted cloud. The white circles mark the BGPS clumps, and each clump name is placed nearby in green text. Bottom panel: The average \co~ and \xco~spectral profiles of the clumps. Each subplot is labelled with its clump ID.}
\label{spec-clumps}
\end{figure}

\begin{figure}[h]
\vspace*{0.2mm}
\centering
\includegraphics[width=0.5\textwidth]
{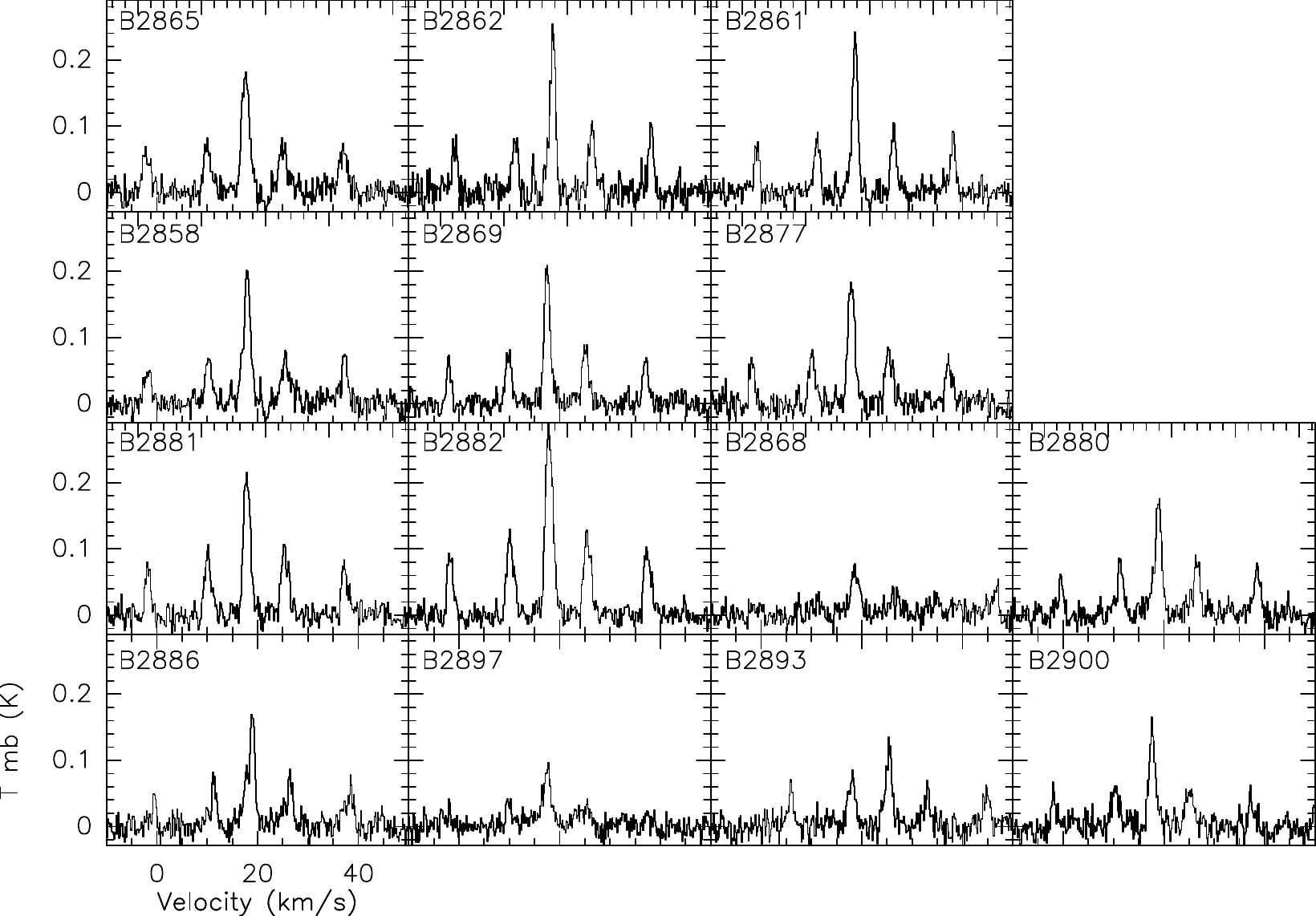}
\caption[]{Average \NH\,(1,1) spectral profile of the clumps. Subplots share the same axis limits.}
\label{spec-clumps_nh11}
\end{figure}

\begin{figure}[h]
\vspace*{0.2mm}
\centering
\includegraphics[width=0.5\textwidth]
{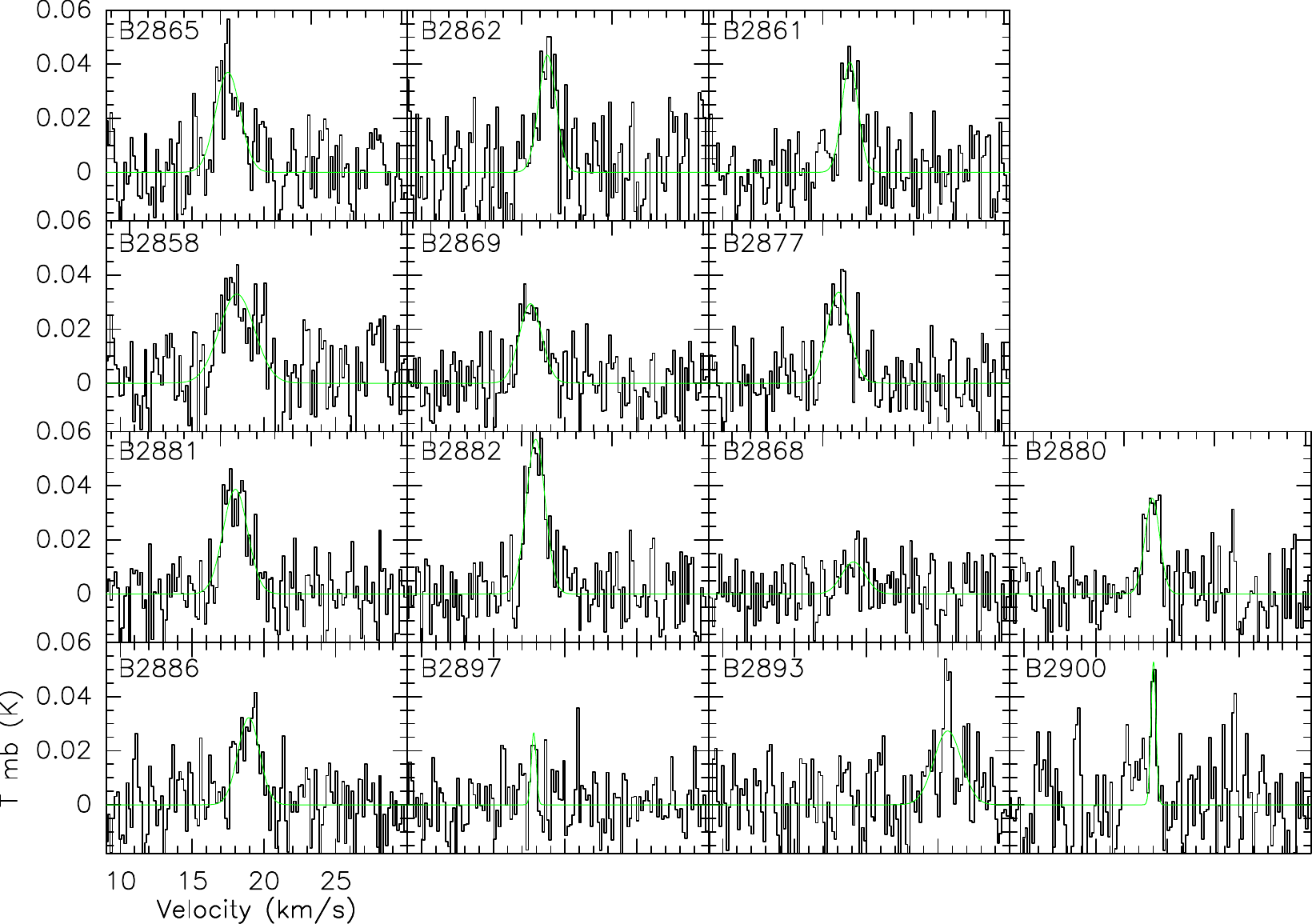}
\caption[]{Average \NH\,(2,2) spectral profile of the clumps. Black lines show the observed spectra, while the green line indicates and 'GAUSS' fitting of NH$_3$\,(2,2) (for illustration only—it does not represent the final fitting result, which was further evaluated). Subplots share the same axis limits.}
\label{spec-clumps_nh22}
\end{figure}

\begin{sidewaystable}[th!]
\setcounter{table}{0}
\caption{Clump properties.}
\label{clumps}
\centering
\small
\begin{tabular}
{p{0.3cm} p{0.7cm} p{0.6cm} p{0.5cm} p{0.5cm} p{1.45cm} p{1.45cm} p{1.3cm} p{1.6cm} p{1.cm} p{1.15cm} p{1.3cm} p{1.3cm} p{1.4cm} p{1.2cm} p{1.cm}}
\hline\hline
Num & Name & Glon & Glat &$R_{\rm clump}$ & $V_{\rm LSR,^{13}CO}$$_{,}$& $\Delta v_{\rm ^{13}CO,}$  & $N_{\rm ^{13}CO}$ & $N_{\rm H_{2}}$($^{13}$CO)$_{,}$ & $M_{\rm clump}$ &   $M_{vir}$ &$T_{\rm kin}$ & $N_{\rm NH_{3}}$ & $\sigma_{\rm NT}$&$\mathcal{M}$ & Profile\\
& & $^{\circ}$ & $^{\circ}$ & pc & $_{\rm clump,}$\kms & $_{\rm clump,}$\kms &  10$^{15}$\,\cmm & $_{\rm clump,}$\,10$^{22}$\,\cmm & \msun & \msun & K & 10$^{14}$\,\cmm &\kms \\
\hline
 1 & B2865 & 13.28 & 0.23  & 0.33 & 17.16 & 0.85(0.01) & 10.30(1.13) & 8.31(0.91) & 197(22)  & 279(32)    & 17.92(0.23) & 3.91(0.56) & 0.29(0.008) & 1.22(0.06) & BA\\
 2 & B2862 & 13.28 & 0.26  & 0.15 & 17.96 & 0.97(0.02) & 9.41(1.37)  & 7.59(1.11) & 36(5)    & 174(27)    & 15.40(0.35) & 2.59(0.36) & 0.21(0.012) & 0.97(0.06) & RP\\
 3 & B2861 & 13.27 & 0.27  & 0.33 & 18.09 & 0.94(0.01) & 9.47(1.33)  & 7.64(1.07) & 181(25)  & 359(46)    & 15.13(0.12) & 2.62(0.35) & 0.21(0.012) & 0.97(0.06) & BW\\
 4 & B2858 & 13.26 & 0.23  & 0.40 & 17.33 & 0.87(0.01) & 9.96(1.22)  & 8.04(0.99) & 285(35)  & 348(50)    & 15.29(0.25) & 1.41(0.22) & 0.22(0.012) & 1.03(0.07) & BA\\
 5 & B2869 & 13.30 & 0.20  & 0.26 & 17.41 & 0.92(0.01) & 8.88(1.17)  & 7.16(0.94) & 103(14)  & 253(29)    & 12.93(0.09) & 3.16(0.67) & 0.31(0.010) & 1.49(0.14) & BA\\
 6 & B2877 & 13.33 & 0.21  & 0.29 & 17.71 & 1.02(0.01) & 11.42(1.51) & 9.21(1.22) & 173(23)  & 364(39)    & 14.35(0.30) & 1.50(0.29) & 0.40(0.007) & 1.85(0.17) & BP\\
 7 & B2881 & 13.34 & 0.21  & 0.29 & 17.97 & 1.03(0.01) & 11.99(1.50) & 9.68(1.21) & 181(23)  & 362(33)    & 15.30(0.14) & 5.10(0.72) & 0.26(0.013) & 1.20(0.10) & BP\\
 8 & B2882 & 13.34 & 0.20  & 0.55 & 17.98 & 1.08(0.01) & 11.32(1.50) & 9.14(1.21) & 602(80)  & 765(75)    & 16.57(0.43) & 5.19(0.64) & 0.27(0.013) & 1.17(0.10) & P\\
 9 & B2868 & 13.29 & 0.15  & 0.70 & 18.49 & 1.26(0.03) & 7.69(1.29)  & 6.20(1.04) & 656(110) & 1259(369)  &             &            &             &            & RP\\
10 & B2880 & 13.34 & 0.16  & 0.59 & 18.61 & 1.07(0.02) & 8.91(1.34)  & 7.19(1.08) & 539(81)  & 788(129)   & 16.46(0.45) & 1.89(0.21) & 0.26(0.011) & 1.16(0.08) & BW\\
11 & B2852 & 13.24 & 0.16  & 0.29 & 18.28 & 1.64(0.07) & 4.97(1.28)  & 4.01(1.03) & 75(19)   & 973(650)   &             &            &             &            & BW\\
12 & B2886 & 13.37 & 0.17  & 0.48 & 18.55 & 0.99(0.01) & 7.87(1.22)  & 6.35(0.98) & 314(49)  & 547(76)    &             &            &             &            & BW\\
13 & B2848 & 13.23 & 0.17  & 0.51 & 18.37 & 1.76(0.07) & 5.58(0.71)  & 4.50(0.57) & 258(33)  & 1805(1263) &             &            &             &            & BW\\
14 & B2897 & 13.43 & 0.25  & 0.59 & 17.56 & 1.14(0.02) & 7.52(1.17)  & 6.07(0.95) & 455(71)  & 914(159)   &             &            &             &            & RA\\
15 & B2841 & 13.19 & 0.17  & 0.22 & 18.81 & 0.94(0.05) & 3.38(0.72)  & 2.73(0.58) & 29(6)    & 232(109)   &             &            &             &            & * \\
16 & B2876 & 13.32 & 0.10  & 0.15 & 18.88 & 0.98(0.06) & 1.98(0.70)  & 1.60(0.57) & 7(3)     & 150(87)    &             &            &             &            & BW*\\
17 & B2864 & 13.28 & 0.10  & 0.40 & 19.62 & 0.83(0.06) & 2.07(0.57)  & 1.67(0.46) & 59 (16)  & 318(184)   &             &            &             &            & RA*\\
18 & B2893 & 13.41 & 0.12  & 0.55 & 18.16 & 0.80(0.02) & 2.58(0.91)  & 2.08(0.73) & 137(48)  & 396(83)    &             &            &             &            & RA*\\
19 & B2900 & 13.47 & 0.18  & 0.40 & 17.85 & 1.00(0.01) & 9.35(1.57)  & 7.55(1.27) & 267(45)  & 479(48)    &             &            &             &            & BW\\
20 & B2831 & 13.15 & 0.10  & 	  &       &            &             &            &            &            &             &            &             &          & *  \\
21 & B2830 & 13.14 & 0.08  & 	  &       &            &             &            &            &            &             &            &             &          & *  \\

\hline
\end{tabular}
\tablefoot{The last column, Profile, represents the line profile characteristics. The abbreviations are as follows: BP (Blue Profile), BW (Blue Wing), RP (Red Profile), RW (Red Wing), BA (Blue Asymmetry), RA (Red Asymmetry), and P (Pedestal).'*' labels the clumps, which may be misclassified or cannot be classified.}
\end{sidewaystable}
\end{appendix}

\end{document}